\documentclass[11pt]{article} 
\usepackage{jheppub}
\usepackage[hhmmss]{datetime}
\usepackage[english]{babel}
\usepackage{amssymb,amsmath,amsfonts,graphicx,mathrsfs,color,bm}
\allowdisplaybreaks  
\usepackage{graphicx}
\usepackage{dcolumn}
\usepackage{hyperref}
\hypersetup{colorlinks=true,linkcolor=magenta,anchorcolor=green,citecolor=cyan,filecolor=black,menucolor=black,urlcolor=cyan}

\usepackage{cancel}
\usepackage{subcaption}
\usepackage{longtable} 
\definecolor{darkred}{rgb}{0.8,0,0}
\definecolor{darkcerulean}{rgb}{0.03, 0.37, 0.69}
\definecolor{vertcem}{cmyk}{0.75 , 0.1 , 0.90, 0}

\newcommand{\be}{\begin{equation}}
\newcommand{\ee}{\end{equation}}
\newcommand{\sbe}{\begin{subequations}}
\newcommand{\see}{\end{subequations}}
\newcommand{\bea}{\begin{eqnarray}}
\newcommand{\eea}{\end{eqnarray}}
\newcommand{\p}{\partial}
\newcommand{\nn}{\nonumber}
\newcommand{\ud}{\mathrm{d}}

\newcommand{\bdm}{\begin{displaymath}}
\newcommand{\edm}{\end{displaymath}}

\usepackage{tikz,fp}
\usetikzlibrary{external,fixedpointarithmetic,decorations.pathmorphing,positioning,calc,fadings,decorations.markings,decorations.pathreplacing,patterns,arrows}
\tikzset{
  label distance=-3pt,
  >=stealth,
  inner sep=1.5pt,
  boson/.style={
    decoration={snake, segment length=2mm, amplitude=0.5mm},
    decorate,
  },
  scalar/.style={
    dashed
  }
}

\title{Scattering angle at 3PM in scalar-tensor theories using the PM-EFT formalism}

\author[a]{Laura Bernard}
 \emailAdd{laura.bernard@obspm.fr}
 \affiliation[a]{%
 Laboratoire d'étude de l'Univers et des phénomènes eXtrêmes (LUX), Observatoire de Paris, Université PSL, Sorbonne Université, CNRS, 92190 Meudon, France
}%

\author[b,c]{Tamanna Jain}%
 \emailAdd{tj317@cam.ac.uk}
\affiliation[b]{LPENS, Département de physique, Ecole normale supérieure, Université PSL, Sorbonne Université, Université Paris Cité, CNRS, 75005 Paris}
    \affiliation[c]{%
 Department of Applied Mathematics and Theoretical Physics,University of Cambridge, Wilberforce Road CB3 0WA Cambridge, United Kingdom
}%

\author[a]{Stavros Mougiakakos}
 \emailAdd{stavros.mougiakakos@obspm.fr}%

\date{\today}

\abstract{
In this work, we derive the conservative dynamics of non-spinning binaries in the massless scalar-tensor theories using the post-Minkowskian Effective Field Theory (EFT) approach. Our main result is an analytic expression of the scattering angle, computed up to third Post-Minkowskian order via two-loop Feynman diagrams. Our results are in perfect agreement with previous literature, in particular within the post-Newtonian limit.
}

\begin{document}
\maketitle
\flushbottom


\section{\label{sec:intro} Introduction}

The advent of gravitational wave astronomy opens a new avenue to deepen our understanding of the gravitational interaction. One way to challenge General Relativity (GR), the current paradigm, is to study theories that go beyond it. One can distinguish two different approaches to test gravity with gravitational wave detectors. First, the agnostic approach, currently being pursued by LIGO-Virgo-KAGRA, searches for parametric deviations from GR~\cite{LIGOScientific:2021sio}. However, such an approach faces several limitations, the main one being that it overlooks physical expectations regarding realistic deviations. As a consequence, the derived constraints are generally not optimal and difficult to interpret in terms of constraints on existing theories of gravity. See Ref~\cite{Bernard:2025dyh} for a recent work that tries to bypass this issue by building a “dictionary” to facilitate the identification of potential deviations from GR in gravitational waveforms based on a local Effective Field Theory (EFT) approach to formulate a departure of GR in terms of curvature operators. The second approach consists in building complete inspiral, merger and ringdown (IMR) waveforms in a specific alternative theory of gravity that can directly be used as an alternative to the GR waveforms in the data analysis of the GW detectors. While it has the advantage of modeling all possible deviations and directly linking to the specific theory in question, it is also a time-demanding task that can reasonably not be pursued for all existing gravity theories. 

The latter theory-specific approach has been extensively followed in the recent years and we are now close to have the first full IMR waveform template beyond GR. The most studied theories lie in the broad class of scalar-tensor theories of gravity. The merger of two binary neutron stars in scalar-tensor theories have been obtained by the pionneered work~\cite{Barausse:2012da,Palenzuela:2013hsa}. More recently, recent advances in the derivation of a well-posed formulation in a large class of Horndeski allowed for the first numerical simulations of the merger of binary systems in Einstein-scalar-Gauss-Bonnet (EsGB) gravity~\cite{East:2020hgw,East:2021bqk, East:2022rqi, AresteSalo:2023mmd}. For the inspiral part, the post-Newtonian formalism~\cite{Blanchet:2013haa} has allowed to get the dynamics at 3PN order~\cite{Bernard:2018hta,Bernard:2018ivi} and the waveform modes at the relative 1.5PN order beyond the GR quadrupolar emission\footnote{This is formally the 2.5PN order beyond the leading dipolar emission of scalar origin.}~\cite{Bernard:2022noq,Jain:2024lie,Trestini:2024mfs,Trestini:2024zpi}. This work has also been completed to include tidal effects~\cite{Bernard:2019yfz,Bernard:2023eul,Dones:2024odv,Dones:2025zbs,Creci:2024wfu}, the leading spin contribution~\cite{Brax:2020vgg,Brax:2021qqo}. It has also been extended to other theories, such as Einstein-Maxwell-dilaton and EsGB~\cite{Julie:2018lfp,Julie:2019sab,Shiralilou:2020gah,Shiralilou:2021mfl,vanGemeren:2023rhh}. Finally, the effective-one-body Hamiltonian and radiation-reaction effects have also been computed to generate waveform templates~\cite{Jain:2022nxs,Jain:2023fvt,Jain:2025sbg,Julie:2022qux,Julie:2024fwy}.

In the present work, we continue in this line of research by computing the scattering angle of two black holes in scalar-tensor theories. To do so, we use the post-Minkowskian Effective Field Theory (PM-EFT) framework~\cite{Kalin:2020mvi,Mogull:2020sak}. Such a framework is complementary to post-Newtonian expansions, as it consists of an expansion in the gravitational constant $G$ only. Results obtained within PM-EFT can then be mapped to bound orbits~\cite{Kalin:2019rwq,Kalin:2019inp,Cho:2021arx} and re-expanded in velocity to compare with traditional PN results and infer higher order PN coefficients.  Current state-of-the-art in GR is the complete 4PM, including radiation-reaction effects, with partial results at 5PM order~\cite{Westpfahl:1985tsl,Damour:2017zjx,Bern:2019nnu,Damour:2019lcq, Kalin:2020fhe,DiVecchia:2021bdo,Bern:2021yeh,Dlapa:2021vgp,Dlapa:2022lmu,Jakobsen:2022fcj,Damgaard:2023ttc,Jakobsen:2023hig,Driesse:2024feo,Driesse:2024xad,Bern:2025wyd}. 

Here, for the first time, we adapt the PM-EFT formalism to scalar-tensor theories, with the goal of reaching the 3PM order by computing two-loop Feynman diagrams. The rest of the paper is organized as follows. In Sec.~\ref{sec:STlagrangian}, we present the scalar-tensor action and express it in a convenient way for EFT calculations. In Sec.~\ref{sec:PMEFT}, we review the PM-EFT formalism and adapt it to scalar-tensor theories by computing the various Feynman rules. Then in Sec.~\ref{sec:consdynamics}, we derive explicitly the conservative dynamics up to 2PM.  Finally, Sec.~\ref{sec:results} presents our main results, namely the impulse and scattering angle up to 3PM. We conclude the paper by comparing our work with PN results in scalar-tensor theories in Sec.~\ref{sec:discuss}. In appendix~\ref{app:Feynman3PM}, we list the different 2-loop Feynman diagrams we have encountered.

Throughout the paper, we use for the metric the mostly minus signature $(+---)$. We also use the shorthand notation $\int_k \equiv \int \frac{d^4k}{(2\pi)^4}$.

\section{\label{sec:STlagrangian} The scalar-tensor Lagrangian}

\subsection{\label{sec:bulkaction} The bulk action}

The bulk action  we will take as a starting point for scalar-tensor theories is the usual Einstein-Hilbert action non-minimally coupled to a massless scalar field $\phi$. It reads
\begin{align}
    S_{\rm bulk} = &\ -\frac{{M_{\rm pl}}^2}{2}\int\ud^{d}x\,\sqrt{-g}\left[\phi R - \frac{\omega(\phi)}{\phi}g^{\alpha\beta}\p_{\alpha}\phi\p_{\beta}\phi\right] 
\end{align}
where $M_{\rm pl}^2=\tfrac{c^3}{8\pi G}\Big(4\pi e^{\gamma_E}L_0^2\Big)^{\frac{4-d}{2}}$, $R$ and $g$ are respectively the Ricci scalar and the determinant of the metric $g^{\alpha \beta}$, $\omega$ is a function of the scalar field.
In order to simplify the calculations, we perform a conformal transformation that will bring the action in another frame, often called the Einstein frame. As it will become evident in the following, our goal is to work with a worldline theory and canonically normalized bulk fields. Specifically, we consider the following conformal transformation for the metric and the redefinition of the scalar field
\begin{align}\label{conftransf}
    &\tilde{g}_{\mu\nu}=\left(\frac{\phi}{\phi_0}\right)^{1-\frac{d-4}{d-2}}g_{\mu\nu}\,,\\ \label{phitransf}
    &\frac{\psi}{\widetilde{M}_{\text{pl}}}=\sqrt{\frac{3+2\omega_0-\frac{d-4}{d-2}}{2}}\ln \frac{\phi}{\phi_0}\,,
\end{align}
where $\phi_0$ is the background asymptotic value of the scalar field, and we have redefined the Planck mass as $\widetilde{M}_{\text{pl}}^2=\phi_0 M_{\rm pl}^2$.
We further expand all the $\phi$-dependent couplings around the asymptotic value $\phi_0$, \textit{e.g.}
\begin{align}
    \omega_0^{(n)}=\frac{d^n\omega(\phi)}{d\phi^n}\Big|_{\phi=\phi_0},\qquad \omega_0^{(0)}\equiv\omega_0\,.
\end{align}
We end up with the following bulk action
\begin{align}\label{SbulkEinst}
    S_{\rm bulk} = &\ \int\ud^{d}x\,\sqrt{-\tilde{g}}\left[-\frac{{\widetilde{M}_{\text{pl}}}^2}{2}\left(\tilde{R}-\frac{1}{2}\tilde{\Gamma}^{\mu}\tilde{\Gamma}_{\mu}\right) +\frac{1}{2}\tilde{g}^{\mu\nu}\partial_{\mu}\psi\partial_{\nu}\psi\sum_{n=0}^{\infty}\frac{c_n}{n!}\left(\frac{\psi}{\widetilde{M}_{\text{pl}}}\right)^n\right]+S_{\rm GF},
\end{align}
where we have added a gauge fixing term $S_{\rm GF}$ in order to work in the harmonic gauge at the level of the action and to simplify the graviton Feynman rules 
 as in~\cite{Kalin:2020mvi}. 
 The coupling constants $c_n$ entering into Eq.~\eqref{SbulkEinst} are defined as
\begin{align}
    c_0=1,\qquad c_n=\left(\frac{2}{3+2\omega_0-\frac{d-4}{d-2}}\right)^{1+n/2}\sum_{m=1}^n(-1)^m\omega_0^{(m)}\phi_0^m\left[\sum_{l=1}^m\frac{l^{n-1}(-1)^l}{\Gamma(l)\Gamma(1+m-l)}\right]\,.
\end{align}

\subsection{\label{sec:matteraction} The matter action}

Following the post-Newtonian EFT approach, we will model the coupling to matter by a worldline action that described the coupling of the two objects to gravity. As is usual in the PM formalism, we start with the quadratic action,
\begin{align}
    S_{\rm mat} = -\sum_{a=1,2}\,\int\ud\sigma_{a}\,\frac{m_{a}\left(\phi\right)}{2}\,e_{a}\left[\frac{1}{e_a^2}\,g_{\mu\nu}\left(x_a^{\alpha}(\sigma)\right)\,v_{a}^{\mu}(\sigma_a)\,v_{a}^{\nu}(\sigma_a)+1\right] + \cdots\,,
\end{align}
where the dots stand for finite-size effects and counter-terms and $v_{a}^{\mu}\equiv\frac{\ud x_{a}^{\mu}}{\ud \sigma}$.

Then, we fix the gauge by imposing $e_a=1$ such that the trajectories are parameterized by the proper time, $\sigma_a = \tau_a$ (contrary to the coordinate time as in PN calculations). It yields the action
\begin{align}
    S_{\rm mat} = -\sum_{a=1,2}\,\int\ud\tau_{a}\,\frac{m_{a}\left(\phi\right)}{2}\,\Bigl[g_{\mu\nu}\left(x_a^{\alpha}(\tau_a)\right)\,v_{a}^{\mu}(\tau_a)\,v_{a}^{\nu}(\tau_a) + 1 \Bigr]\,,
\end{align}
with the condition
\begin{align}
\label{eq:einbein}
    e_a^2 = g_{\mu\nu}\left(x_a^{\alpha}(\tau_a)\right)\,v_{a}^{\mu}(\tau_a)\,v_{a}^{\nu}(\tau_a) = 1\,.
\end{align}
As previously, we apply the conformal transformation to the matter action and we get
\begin{align}\label{Smatter}
    S_{\rm mat} = -\sum_{a=1,2}\,\frac{\overline{m}_{a}}{2}\,\int\ud\tau_{a}\,\left[\tilde{g}_{\mu\nu}\left(x_a^{\alpha}(\tau_a)\right)\,v_{a}^{\mu}(\tau_a)\,v_{a}^{\nu}(\tau_a)\sum_{n=0}^{\infty}\frac{\tilde{d}_n^{(a)}}{n!}\left(\frac{\psi}{\widetilde{M}_{\text{pl}}}\right)^n + \sum_{n=0}^{\infty}\frac{\tilde{f}_n^{(a)}}{n!}\left(\frac{\psi}{\widetilde{M}_{\text{pl}}}\right)^n \right]\,,
\end{align}
where we have defined
\begin{align}
& \overline{m}_a \equiv m_a(\phi_0) \,,\qquad s_a^{(n)}=\frac{d^{n+1}\ln\ m_a(\phi)}{d \ln\phi^{n+1}}\Big|_{\phi=\phi_0}\,.
\end{align}
The coupling constants up to the order of interest for this work are given by 
\begin{align}
\tilde{d}_n^{(a)} &= d_n^{(a)}\left(\frac{2}{3+2\omega_0-\frac{d-4}{d-2}}\right)^{n/2},\nn\\
d_0^{(a)} &= 1,\nn\\
d_1^{(a)} &= s_a^{(0)}-1+\frac{d-4}{d-2},\nn\\
d_2^{(a)} &= s_a^{(1)} + (s_a^{(0)})^2 - 2s_a^{(0)} + 1
            +2\frac{d-4}{d-2}\left(s_a^{(0)}-\frac{d}{2(d-2)}\right),\nn\\
d_3^{(a)} &= s_a^{(2)} - 3s_a^{(1)} +3s_a^{(0)}s_a^{(1)} + (s_a^{(0)})^3
            -3(s_a^{(0)})^2 + 3s_a^{(0)} - 1\\
          &\quad
            +3\frac{d-4}{d-2}\left(s_a^{(1)}+\left(s_a^{(0)}-\frac{d}{2(d-2)}\right)^2
            +\frac{(d-4)^2}{12(d-2)^2}\right)~,
\label{eq:coeff11}            
\end{align}
and
\begin{align}
\label{eq:coeff12}
\tilde{f}_n^{(a)} &= f_n^{(a)}\left(\frac{2}{3+2\omega_0-\frac{d-4}{d-2}}\right)^{n/2},\nn\\
f_0^{(a)} &= 1,\nn\\
f_1^{(a)} &= s_a^{(0)},\nn\\
f_2^{(a)} &= (s_a^{(0)})^2 + s_a^{(1)},\nn\\
f_3^{(a)} &= s_a^{(2)} + 3s_a^{(0)}s_a^{(1)} + (s_a^{(0)})^3~.    
\end{align}

\section{\label{sec:PMEFT} The PM-EFT formalism}
In this section, we will briefly describe the PM-EFT formalism used to obtain the impulse and the scattering angle. We first provide a set of Feynman rules for the massless scalar-tensor theory bulk action and the worldline action up to 3PM order, and then we discuss the computation of the impulse from the effective action. 

\subsection{\label{sec:} Propagator and vertices}
Here, we derive Feynman rules in scalar-tensor gravity theories with the bulk action,
\begin{align}
\label{eq:bulk}
S_{\rm bulk} = \int d^dx \sqrt{-\tilde{g}}\left[-\frac{\widetilde{M}^2_{\text{pl}}}{2}\tilde{R} + \frac{1}{2}\tilde{g}^{\mu \nu}\partial_{\mu}\psi\partial_{\nu}\psi\sum_{n=0}\frac{c_n}{n!}\left(\frac{\psi}{\widetilde{M}_{\text{pl}}}\right)^n\right]+S_{\rm GF}~,
\end{align}
and the point particle action $S_{\rm mat}$ given in Eq.~\eqref{Smatter}.

Considering the expansion of the metric in the weak field approximation, 
\begin{align}
\tilde{g}_{\mu \nu} = \eta_{\mu \nu} + \kappa  h_{\mu \nu}~,
\end{align}
where $\kappa =  \frac{2}{\widetilde{M}_{\text{pl}}}$ is the expansion parameter. In this weak field approximation, the matter action of Eq.~\eqref{Smatter} generates only a single one-point graviton worldline vertex in the pure gravitational-sector, while the scalar sector generates both a one-point scalar worldline vertex and higher-order multi‑leg scalar-graviton worldline vertices. The latter contributions in the scalar sector are due to worldline non-linearities. For the bulk vertices, the action \eqref{eq:bulk} generates the three- and four-graviton vertices up to 3PM order \cite{Kalin:2020mvi}, and in the scalar sector it gives both the pure three- and four-scalar vertices together with the mixed graviton-scalar vertices. The graviton-scalar bulk vertices up to 3PM include 2 scalar-1 graviton, 2 scalar-2 gravitons, and 3 scalar-1 graviton vertices. 

The Feynman rules for all the topologies up to 3PM in the momentum space are given in Table~\ref{tab:Feynman} with 
the following definitions of $\mathcal{P}_{\mu \nu \alpha \beta}$, $\bar{\mathcal{P}}_{\mu \nu \alpha \beta}$, and  $\bar{\mathcal{I}}^{\alpha \beta \mu \nu}{}_{\rho \lambda}$, 
\begin{align}
\label{eq:PGR}
\mathcal{P}_{\mu \nu \alpha \beta} = \frac{1}{2}\left(\eta_{\mu \alpha}\eta_{\nu \beta}+\eta_{\nu \alpha}\eta_{\mu \beta}-\frac{2}{d-2}\eta_{\mu \nu}\eta_{\alpha \beta}\right)~,
\end{align}
\begin{align}
\label{eq:Pscalar}
\bar{\mathcal{P}}_{\mu \nu \alpha \beta} = \frac{1}{2}\left(\eta_{\mu \alpha}\eta_{\nu \beta}+\eta_{\nu \alpha}\eta_{\mu \beta}-\eta_{\mu \nu}\eta_{\alpha \beta}\right)~,
\end{align}
and
\begin{align}
    \bar{\mathcal{I}}^{\alpha \beta \mu \nu}{}_{\rho \lambda}=\Big(\frac{1}{2}\bar{\mathcal{P}}^{\alpha \beta \mu \nu} \eta_{\rho \lambda}-\bar{\mathcal{P}}^{\alpha \beta \mu}{}_{\rho} \delta^{\nu}_{\lambda}-\bar{\mathcal{P}}^{\alpha}{}_{\rho}{}^{\mu \nu} \delta^{\beta}_ {\lambda}\Big).
\end{align}

\setlength{\tabcolsep}{6pt}
\renewcommand{\arraystretch}{1.2}

\begin{longtable}{@{\hspace{1.8em}}p{0.22\textwidth} p{0.30\textwidth} p{0.35\textwidth}@{}}
\caption{Feynman rules for propagators, worldline vertices and bulk vertices in the massless scalar–tensor theory up to 3PM order. Note that the overall normalisation in GR is with respect to $M_{\text{pl}}^{GR} = 1/\sqrt{32\pi G}$ ~\cite{Kalin:2020mvi,Kalin:2020fhe}, whereas in our work the normalisation is with respect to $\widetilde{M}_{\text{pl}}$. The two are related by $\widetilde{M}_{\text{pl}} = 2\phi_0 M_{\text{pl}}^{GR}$.}
\label{tab:Feynman}\\
\hline\hline
Diagram & Notation & Expression \\
\hline\hline
\endfirsthead

\endhead

\hline\hline
\endfoot

\hline\hline
\endlastfoot

\multicolumn{3}{c}{Propagators} \\
\hline
\hspace{0.9em}\raisebox{-0.8\height}{\begin{tikzpicture}[scale=1,>=stealth]
   \draw[boson] (0,0.4) -- (0,-0.4);
   \draw[->] (0.25,-0.2) -- (0.25,0.2);
   \node[right] at (0.25,0.0) {$k$};
\end{tikzpicture}}
&
\raisebox{-1.9\height}{$\langle h_{\mu \nu} (x) h_{\alpha \beta}(y)\rangle$}
&
\raisebox{-1.2\height}{$\displaystyle \frac{i\,\mathcal{P}_{\mu\nu\alpha\beta}(k)}{k^2 } e^{i k\cdot(x-y)}$}
\\
\hspace{0.9em}\raisebox{-0.4\height}{\begin{tikzpicture}[scale=1,>=stealth]
   \draw[scalar] (0,0.4) -- (0.0,-0.4);
   \draw[->] (0.25,-0.2) -- (0.25,0.2);
   \node[right] at (0.25,0.0) {$k$};
\end{tikzpicture}}
&
$\langle \psi (x) \psi(y)\rangle$
&
$\displaystyle \frac{i}{k^2}e^{i k\cdot(x-y)}$
\\
\\
\hline
\multicolumn{3}{c}{Worldline vertices} \\
\hline
\hspace{0.9em}\raisebox{-0.9\height}{\begin{tikzpicture}[scale=1]
   \draw[boson] (0,0.4) -- (0,-0.4);
   \filldraw (0,0.4) circle (0.05);
\end{tikzpicture}}
&
&
\raisebox{-0.7\height}{$\displaystyle -i\frac{ \bar{m}_a}{\widetilde{M}_{\text {pl}}} v_a^{\mu}v_a^{\nu}$}
\\
\hspace{0.9em}\raisebox{-0.9\height}{\begin{tikzpicture}[scale=1]
   \draw[scalar] (0,0.4) -- (0,-0.4);
   \filldraw (0,0.4) circle (0.05);
\end{tikzpicture}}
&
&
\raisebox{-0.7\height}{$\displaystyle -i\frac{ \bar{m}_a}{2 \widetilde{M}_{\text {pl}} } \Big(\tilde{d}_{1}^{(a)}v_a^{\mu}v_{a,\mu} + \tilde{f}_{1}^{(a)} \Big)$}
\\
\raisebox{-0.9\height}{\begin{tikzpicture}[scale=1]
   \draw[boson] (0,0.4) -- (-0.4,-0.4);
   \draw[scalar] (0,0.4) -- (0.4,-0.4);
   \filldraw (0,0.4) circle (0.05);
\end{tikzpicture}}
&
&
\raisebox{-0.7\height}{$\displaystyle -i\frac{ \bar{m}_a \tilde{d}_1^{(a)}}{\widetilde{M}_{\text{pl}}^2} v_a^{\mu}v_a^{\nu}$}
\\
\raisebox{-0.9\height}{\begin{tikzpicture}[scale=1]
   \draw[scalar] (0,0.4) -- (-0.4,-0.4);
   \draw[scalar] (0,0.4) -- (0.4,-0.4);
   \filldraw (0,0.4) circle (0.05);
\end{tikzpicture}}
&
&
\raisebox{-0.7\height}{$\displaystyle -i\frac{\bar{m}_a}{4\widetilde{M}_{\text{pl}}^2} \Big(\tilde{d}_2^{(a)}v_a^{\mu}v_{a,\mu}  + \tilde{f}_2^{(a)}\Big)$}
\\
\raisebox{-0.9\height}{\begin{tikzpicture}[scale=1]
   \draw[scalar] (0,0.4) -- (-0.4,-0.4);
   \draw[scalar] (0,0.4) -- (0.0,-0.4);
   \draw[boson] (0,0.4) -- (0.4,-0.4);
   \filldraw (0,0.4) circle (0.05);
\end{tikzpicture}}
&
&
\raisebox{-0.7\height}{$\displaystyle -i\frac{\bar{m}_a}{2\widetilde{M}_{\text{pl}}^3} \Big(\tilde{d}_2^{(a)}v_a^{\mu}v_{a}^{\nu}\Big)$}
\\
\raisebox{-0.9\height}{\begin{tikzpicture}[scale=1]
   \draw[scalar] (0,0.4) -- (-0.4,-0.4);
   \draw[scalar] (0,0.4) -- (0.0,-0.4);
   \draw[scalar] (0,0.4) -- (0.4,-0.4);
   \filldraw (0,0.4) circle (0.05);
\end{tikzpicture}}
&
&
\raisebox{-0.7\height}{$\displaystyle -i\frac{\bar{m}_a}{2\widetilde{M}_{\text{pl}}^3} \frac{1}{3!}\Big(\tilde{d}_3^{(a)}v_a^{\mu}v_{a,\mu}  + \tilde{f}_3^{(a)}\Big)$}
\\
\\
\hline
\multicolumn{3}{c}{Bulk vertices} \\
\hline
\raisebox{-0.9\height}{\begin{tikzpicture}[scale=1,>=stealth]
  \filldraw (0,0) circle (0.06);
  \draw[boson] (0,0) -- (0,0.6);
  \draw[->] (0.15,0.2) -- (0.15,0.6);
  \node[right] at (0.20,0.55) {\scriptsize$k_{1}$};
  \draw[boson] (0,0) -- (-0.6,-0.6);
  \draw[->] (-0.35,-0.2) -- (-0.70,-0.55);
  \node[left] at (-0.65,-0.45) {\scriptsize$k_{2}$};
  \draw[boson] (0,0) -- (0.6,-0.6);
  \draw[->] (0.35,-0.2) -- (0.70,-0.55);
  \node[right] at (0.65,-0.45) {\scriptsize$k_{3}$};
\end{tikzpicture}
}
&
\raisebox{-1.2\height}{$i V_{hhh}^{\mu \nu \alpha \beta \rho \lambda}(k_1,k_2,k_3)$}
&
\raisebox{-1.2\height}{\hspace{0.9em} Eq.~(4.4) of Ref.~\cite{Kalin:2020mvi}}
\\
\raisebox{-0.9\height}{\begin{tikzpicture}[scale=1,>=stealth]
  \filldraw (0,0) circle (0.06);
  \draw[scalar] (0,0) -- (0,0.6);
  \draw[->] (0.15,0.2) -- (0.15,0.6);
  \node[right] at (0.20,0.55) {\scriptsize$k_{1}$};
  \draw[scalar] (0,0) -- (-0.6,-0.6);
  \draw[->] (-0.35,-0.2) -- (-0.70,-0.55);
  \node[left] at (-0.65,-0.45) {\scriptsize$k_{2}$};
  \draw[scalar] (0,0) -- (0.6,-0.6);
  \draw[->] (0.35,-0.2) -- (0.70,-0.55);
  \node[right] at (0.65,-0.45) {\scriptsize$k_{3}$};
\end{tikzpicture}}
&
\raisebox{-1.2\height}{$i V_{\psi \psi \psi} (k_1,k_2,k_3)$}
&
\hspace{0.9em}\raisebox{-1.0\height}{$\displaystyle i\frac{c_1}{2\widetilde{M}_{\text{pl}}}\Big(k_1^2+k_2^2+k_3^2\Big)$}
\\
\raisebox{-0.9\height}{\begin{tikzpicture}[scale=1,>=stealth]
  \filldraw (0,0) circle (0.06);
  \draw[scalar] (0,0) -- (0,0.6);
  \draw[->] (0.15,0.2) -- (0.15,0.6);
  \node[right] at (0.20,0.55) {\scriptsize$k_{1}$};
  \draw[scalar] (0,0) -- (-0.6,-0.6);
  \draw[->] (-0.35,-0.2) -- (-0.70,-0.55);
  \node[left] at (-0.65,-0.45) {\scriptsize$k_{2}$};
  \draw[boson] (0,0) -- (0.6,-0.6);
  \draw[->] (0.35,-0.2) -- (0.70,-0.55);
  \node[right] at (0.65,-0.45) {\scriptsize$k_{3}$};
\end{tikzpicture}}
&
\raisebox{-1.2\height}{$ i V_{\psi \psi h}^{\alpha \beta} (k_1,k_2)$}
&
\hspace{0.9em}\raisebox{-1.0\height}{$\displaystyle i\frac{2}{\widetilde{M}_{\text{pl}}}\bar{\mathcal{P}}^{\alpha \beta}_{\mu \nu} k_1^{\mu} k_2^{\nu}$}
\\
\raisebox{-0.9\height}{\begin{tikzpicture}[scale=1,>=stealth]
  \filldraw (0,0) circle (0.06);
  \draw[scalar] (0,0) -- (0.6, 0.6);
  \draw[->] (0.35,0.2) -- (0.70,0.55);
  \node[right] at (0.65,0.45) {\scriptsize$k_{1}$};
\draw[scalar] (0,0) -- (-0.6,0.6);
  \draw[->] (-0.35,0.2) -- (-0.70,0.55);
  \node[right] at (-1.1,0.45) {\scriptsize$k_{2}$};
  \draw[scalar] (0,0) -- (-0.6,-0.6);
  \draw[->] (-0.35,-0.2) -- (-0.70,-0.55);
  \node[left] at (-0.65,-0.45) {\scriptsize$k_{3}$};
  \draw[scalar] (0,0) -- (0.6,-0.6);
  \draw[->] (0.35,-0.2) -- (0.70,-0.55);
  \node[right] at (0.65,-0.45) {\scriptsize$k_{4}$};
\end{tikzpicture}}
&  \raisebox{-1.2\height}{$i V_{\psi \psi \psi \psi} (k_1,k_2,k_3,k_4)$}
& \raisebox{-1.0\height}{$\displaystyle \,-i\frac{c_2}{2 \widetilde{M}_{\text{pl}}^2} \sum_{1\leq i<j\leq4} k_i \cdot k_j$}
\\
\raisebox{-0.9\height}{\begin{tikzpicture}[scale=1,>=stealth]
  \filldraw (0,0) circle (0.06);
  \draw[scalar] (0,0) -- (0.6, 0.6);
  \draw[->] (0.35,0.2) -- (0.70,0.55);
  \node[right] at (0.65,0.45) {\scriptsize$k_{1}$};
\draw[scalar] (0,0) -- (-0.6,0.6);
  \draw[->] (-0.35,0.2) -- (-0.70,0.55);
  \node[right] at (-1.1,0.45) {\scriptsize$k_{2}$};
  \draw[scalar] (0,0) -- (-0.6,-0.6);
  \draw[->] (-0.35,-0.2) -- (-0.70,-0.55);
  \node[left] at (-0.65,-0.45) {\scriptsize$k_{3}$};
  \draw[boson] (0,0) -- (0.6,-0.6);
  \draw[->] (0.35,-0.2) -- (0.70,-0.55);
  \node[right] at (0.65,-0.45) {\scriptsize$k_{4}$};
\end{tikzpicture}}
& \raisebox{-1.3\height}{$i V_{\psi \psi \psi h}^{\alpha \beta} (k_1,k_2,k_3)$}
& \hspace{0.9em}\raisebox{-1.0\height}{$\displaystyle \,i\frac{2c_1}{\widetilde{M}_{\text{pl}}^2}\bar{\mathcal{P}}_{\mu \nu}^{\alpha \beta} \big(k_1^{\mu }k_2^{\nu}+k_2^{\mu }k_3^{\nu}+k_1^{\mu }k_3^{\nu}\big)$}
\\
\raisebox{-0.9\height}{\begin{tikzpicture}[scale=1,>=stealth]
  \filldraw (0,0) circle (0.06);
  \draw[scalar] (0,0) -- (0.6, 0.6);
  \draw[->] (0.35,0.2) -- (0.70,0.55);
  \node[right] at (0.65,0.45) {\scriptsize$k_{1}$};
\draw[scalar] (0,0) -- (-0.6,0.6);
  \draw[->] (-0.35,0.2) -- (-0.70,0.55);
  \node[right] at (-1.1,0.45) {\scriptsize$k_{2}$};
  \draw[boson] (0,0) -- (-0.6,-0.6);
  \draw[->] (-0.35,-0.2) -- (-0.70,-0.55);
  \node[left] at (-0.65,-0.45) {\scriptsize$k_{3}$};
  \draw[boson] (0,0) -- (0.6,-0.6);
  \draw[->] (0.35,-0.2) -- (0.70,-0.55);
  \node[right] at (0.65,-0.45) {\scriptsize$k_{4}$};
\end{tikzpicture}}
&  \raisebox{-1.2\height}{$i V_{\psi \psi h h}^{\alpha \beta \mu \nu} (k_1,k_2)$}
& \hspace{0.9em}\raisebox{-0.9\height}{$i\frac{4 }{\widetilde{M_{\rm pl}^2}} \bar{\mathcal{I}}^{\alpha \beta \mu \nu}{}_{\rho \lambda} k_1^{\rho}k_2^{\lambda}$}
\\
\raisebox{-0.9\height}{\begin{tikzpicture}[scale=1,>=stealth]
  \filldraw (0,0) circle (0.06);
  \draw[boson] (0,0) -- (0.6, 0.6);
  \draw[->] (0.35,0.2) -- (0.70,0.55);
  \node[right] at (0.65,0.45) {\scriptsize$k_{1}$};
\draw[boson] (0,0) -- (-0.6,0.6);
  \draw[->] (-0.35,0.2) -- (-0.70,0.55);
  \node[right] at (-1.1,0.45) {\scriptsize$k_{2}$};
  \draw[boson] (0,0) -- (-0.6,-0.6);
  \draw[->] (-0.35,-0.2) -- (-0.70,-0.55);
  \node[left] at (-0.65,-0.45) {\scriptsize$k_{3}$};
  \draw[boson] (0,0) -- (0.6,-0.6);
  \draw[->] (0.35,-0.2) -- (0.70,-0.55);
  \node[right] at (0.65,-0.45) {\scriptsize$k_{4}$};
\end{tikzpicture}}
& \raisebox{-1.2\height}{$i V_{hhhh}^{\mu \nu \alpha \beta \rho \lambda \gamma \delta} (k_1,k_2,k_3,k_4)$}
& \hspace{0.9em}\raisebox{-1.0\height}{Ref.~\cite{Kalin:2020fhe}}
\\
\\
\end{longtable}

\subsection{\label{sec:impulse} Impulse}

Given the matter action~\eqref{Smatter} and bulk action~\eqref{SbulkEinst}, the effective two-body action $S_{\rm eff}$ in scalar-tensor theories is constructed by integrating out the graviton and scalar fields,
\begin{align}
    e^{i S_{\rm eff}[x_a]} = \int \mathcal{D}\psi\mathcal{D}h_{\mu \nu}  e^{i S_{\rm bulk}[h,\psi]+ iS_{\rm mat}[h,x_a]}~.
\end{align}
The zeroth order contribution to the effective action obtained by integrating out the fields around the background field $\eta_{\mu\nu}$ is \cite{Damour:1995kt},
    \begin{align}
    \label{eq:SeffLO}
    e^{i S_{\rm eff}[x_a]} = e^{\left(-\sum_{a=1,2}i\frac{\bar{m}_a}{2}\int d{\tau_a}\left[\eta_{\mu \nu} v^{\mu}_a v^\nu_a\tilde{d}_0^{(a)}+\tilde{f}_0^{(a)}\right]+iB_{\rm bulk}^{hh}+iB_{\rm bulk}^{\psi \psi}\right)}~,
    \end{align}
    where $B^{hh}_{\rm bulk}$ and $B^{\psi \psi}_{\rm bulk}$ are the tree-level bulk contribution to the effective action due to graviton and scalar exchange, respectively.

Using this effective action, we can derive the equations of motion as a perturbative series which then can be used to compute the impulse, and hence the scattering angle. Without loss of generality, let us focus on the deflection for particle 1. In the PM expansion, the effective action takes the form,
    \begin{align}
        \label{eq:SeffLO1}
        S_{\rm eff} = \sum_{n}\int d{\tau_1} \mathcal{L}_n[x_1(\tau_1),x_2(\tau_2)]~,
    \end{align}
where $n$ denotes the order of the PM expansion and $\mathcal{L}_n$'s are $\mathcal{O} (G^n)$ contribution to the effective Lagrangian obtained by summing all the $n$-PM Feynman diagrams.  

From Eqs.~\eqref{eq:SeffLO} and \eqref{eq:SeffLO1}, the zeroth order contribution to the Lagrangian $\mathcal{L}_0$ for particle 1 becomes,
    \begin{align}
        \mathcal{L}_0 & = -\int d{\tau_1}\frac{\bar{m}_1}{2}\left[\eta_{\mu \nu} v^{\mu}_1 v^\nu_1\tilde{d}_0^{(1)}+\tilde{f}_0^{(1)}\right]\nn \\
         & = -\int d\tau_1 \frac{\bar{m}_1}{2} \eta_{\mu \nu} v^{\mu}_1 v^\nu_1~.
    \end{align}
The leading order (kinetic) term in the scalar-tensor theories coincides with the pure general relativity result (see Eq.~(3.20) of Ref.~\cite{Kalin:2020mvi}). Therefore, the impulse and scattering angle in scalar-tensor theories can be computed by exactly following the procedure of~\cite{Kalin:2020mvi}, now with the corrections due to scalar exchanges. 

The total change of the four-momentum for particle 1 is then,
\begin{align}
    \Delta p_1^{\mu} = \bar{m}_1 \Delta v_1^{\mu} = -\eta^{\mu \nu} \sum_{n}\int_{-\infty}^{\infty} d \tau_1 \frac{\partial \mathcal{L}_n}{\partial x_1^{\nu}}~,
\end{align}
where the impulse is computed by including the deflected worldline trajectories. By expanding the dynamics order by order in the PM expansion series, the impulse at the $n$-th order is,
\begin{align}
    \Delta^{(n)} p^{\mu}_a & = \sum_{k\leq n} \Delta_{\mathcal{L}_{k}}^{(n)} p^{\mu}_a~,
\end{align}
where the $\Delta_{\mathcal{L}_{k}}$ denotes the contribution due to the $k$-PM Lagrangian. Explicitly,
\begin{align}
    \Delta_{\mathcal{L}_{k}}^{(n)} p^{\mu}_a & = -\eta^{\mu \nu} \int_{-\infty}^{\infty} d\tau_a \Big(\frac{\partial}{\partial x^{\nu}_a} \mathcal{L}_k [b_a + u_a \tau_a + \sum_{r=0}^{n-k}\delta^{(r)}x_a]\Big)_{\mathcal{O}(G)^{n}}~,
\end{align}
where we have used the PM-deflected trajectory $x_a^{\mu} = b_a^{\mu} + u_a^{\mu} \tau_a + \sum_n \delta^{(n)}x_a^{\mu}$ in $\mathcal{L}_{k}$, expanded in powers of $G$ and retained terms only up to the order of $\mathcal{O}(G)^{n}$ (see Ref.~\cite{Kalin:2020mvi} for more details). Here, $u_a$ is the initial velocity at infinity, $b_a\cdot u_a=0 $ and $b^{\mu}\equiv b_1^{\mu}-b_2^{\mu}$ is associated with the impact parameter of the collision in the center-of-mass frame. 

\section{\label{sec:consdynamics} Conservative Dynamics}
In this section, we compute the effective action and trajectories using the EFT approach. We explicitly illustrate all the steps up to 2PM order, starting from the computation of the effective Lagrangian following the same approach as in pure GR, by integrating out both the metric and scalar degrees of freedom, but keeping the propagators fully relativistic. From the Lagrangian, we will compute the equations of motion, hence, the corrections to the undeflected trajectories. 

The entire set of diagrams that contribute to the effective action up to 2PM order are given in Fig.~\ref{diag:Feynman2PM} and the entire set of diagrams required at 3PM order are given in Appendix~\ref{app:Feynman3PM}. 

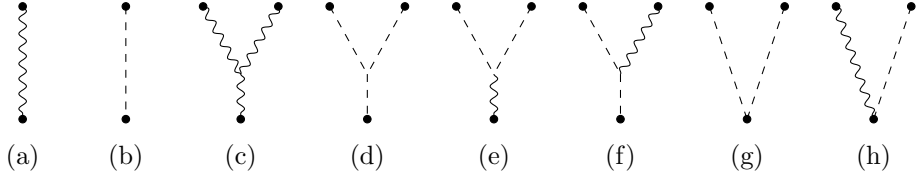
\begin{figure}[h]
  \centering
  \begin{subfigure}[b]{0.08\textwidth}
    \centering
    \begin{tikzpicture}
      [scale=1]
      \draw[boson] (0,0) -- (0,-1.5);
      \filldraw (0,0) circle (0.05);
      \filldraw (0,-1.5) circle (0.05);
    \end{tikzpicture}
    \caption{}
  \end{subfigure}
    \begin{subfigure}[b]{0.08\textwidth}
    \centering
    \begin{tikzpicture}
      [scale=1]
      \draw[scalar] (0,0) -- (0,-1.5);
      \filldraw (0,0) circle (0.05);
      \filldraw (0,-1.5) circle (0.05);
    \end{tikzpicture}
    \caption{}
  \end{subfigure}
    \begin{subfigure}[b]{0.1\textwidth}
    \centering
    \begin{tikzpicture}
      [scale=1]
      \draw[boson] (-0.5,0) -- (0,-0.9);
      \draw[boson] (0.5,0) -- (0,-0.9);
      \draw[boson] (0,-0.9) -- (0,-1.5);
      \filldraw (-0.5,0) circle (0.05);
      \filldraw (0.5,0) circle (0.05);
      \filldraw (0,-1.5) circle (0.05);
    \end{tikzpicture}
    \caption{}
  \end{subfigure}
    \begin{subfigure}[b]{0.1\textwidth}
    \centering
    \begin{tikzpicture}
      [scale=1]
      \draw[scalar] (-0.5,0) -- (0,-0.9);
      \draw[scalar] (0.5,0) -- (0,-0.9);
      \draw[scalar] (0,-0.9) -- (0,-1.5);
      \filldraw (-0.5,0) circle (0.05);
      \filldraw (0.5,0) circle (0.05);
      \filldraw (0,-1.5) circle (0.05);
    \end{tikzpicture}
    \caption{}
  \end{subfigure}
    \begin{subfigure}[b]{0.1\textwidth}
    \centering
    \begin{tikzpicture}
      [scale=1]
      \draw[scalar] (-0.5,0) -- (0,-0.9);
      \draw[scalar] (0.5,0) -- (0,-0.9);
      \draw[boson] (0,-0.9) -- (0,-1.5);
      \filldraw (-0.5,0) circle (0.05);
      \filldraw (0.5,0) circle (0.05);
      \filldraw (0,-1.5) circle (0.05);
    \end{tikzpicture}
    \caption{}
  \end{subfigure}
    \begin{subfigure}[b]{0.1\textwidth}
    \centering
    \begin{tikzpicture}
      [scale=1]
      \draw[scalar] (-0.5,0) -- (0,-0.9);
      \draw[boson] (0.5,0) -- (0,-0.9);
      \draw[scalar] (0,-0.9) -- (0,-1.5);
      \filldraw (-0.5,0) circle (0.05);
      \filldraw (0.5,0) circle (0.05);
      \filldraw (0,-1.5) circle (0.05);
    \end{tikzpicture}
    \caption{}
  \end{subfigure}
        \begin{subfigure}[b]{0.1\textwidth}
    \centering
    \begin{tikzpicture}
      [scale=1]
      \draw[scalar] (-0.5,0) -- (0,-1.5);
      \draw[scalar] (0.5,0) -- (0,-1.5);
      \filldraw (-0.5,0) circle (0.05);
      \filldraw (0.5,0) circle (0.05);
      \filldraw (0,-1.5) circle (0.05);
    \end{tikzpicture}
    \caption{}
  \end{subfigure}
        \begin{subfigure}[b]{0.1\textwidth}
    \centering
    \begin{tikzpicture}
      [scale=1]
      \draw[boson] (-0.5,0) -- (0,-1.5);
      \draw[scalar] (0.5,0) -- (0,-1.5);
      \filldraw (-0.5,0) circle (0.05);
      \filldraw (0.5,0) circle (0.05);
      \filldraw (0,-1.5) circle (0.05);
    \end{tikzpicture}
    \caption{}
  \end{subfigure}
    \caption{Feynman diagrams needed for the computation of the effective action up to 2PM order. The wavy lines represent the graviton propagator, dashed lines represents the scalar propagator and the black dots represent the external sources.}
    \label{diag:Feynman2PM}
\end{figure}  

\subsection{Tree level}

The effective Lagrangian at the leading order (LO) due to the two `tree'-diagrams on the left-hand-side of Fig.~\ref{diag:Feynman2PM}, displaying respectively one scalar and one graviton exchange, is the following,

\begin{align}
\mathcal{L}_1 =& -i \left(\frac{-i \bar{m}_1}{\widetilde{M}_{\rm pl}}\right)\left(\frac{-i \bar{m}_2} {\widetilde{M}_{\rm pl}}\right)\int d\tau_2v_2^{\alpha}(\tau_2)v_2^{\beta}(\tau_2)v_1^{\mu}(\tau_1)v_1^{\nu}(\tau_1)\int_k i \frac{\mathcal{P}_{\mu \nu \alpha \beta}}{k^2}e^{i k(x_1(\tau_1)-x_1(\tau_2))}\nonumber \\
&-i\left(\frac{-i \bar{m}_1}{2\widetilde{M}_{\rm pl}}\right)\left(\frac{-i \bar{m}_2} {2\widetilde{M}_{\rm pl}}\right)\int d\tau_2  \left(\tilde{d}_1^{(1)}v_1^2(\tau_1)+\tilde{f}_1^{(1)}\right)\left(\tilde{d}_1^{(2)}v_2^2(\tau_2)+\tilde{f}_1^{(2)}\right)\int_k \frac{i}{k^2}e^{i k(x_1(\tau_1)-x_1(\tau_2))}~.
\end{align}
Inserting $\mathcal{P}_{\mu \nu \alpha \beta}$ from Eq.~\eqref{eq:PGR}, the LO Lagrangian in $d$-dimensions is then,
\begin{align}
\label{1PML}
\mathcal{L}_1 =& -\frac{ \bar{m}_1\bar{m}_2}{2\widetilde{M}_{\rm pl}^2}\int d\tau_2\left(2(v_2(\tau_2)\cdot v_1(\tau_1))^2-\frac{2}{d-2}v_1^2(\tau_1)v_2^2(\tau_2)\right)\int_k  \frac{1}{k^2}e^{i k\cdot(x_1(\tau_1)-x_1(\tau_2))}\nonumber \\
&-\frac{ \bar{m}_1\bar{m}_2}{4\widetilde{M}_{\rm pl}^2}\int d\tau_2  \left(\tilde{d}_1^{(1)}v_1^2(\tau_1)+\tilde{f}_1^{(1)}\right)\left(\tilde{d}_1^{(2)}v_2^2(\tau_2)+\tilde{f}_1^{(2)}\right)\int_k \frac{1}{k^2}e^{i k\cdot(x_1(\tau_1)-x_1(\tau_2))}~.
\end{align}
Then, the contributions to the equations of motion (EOM), 
\begin{align}
\label{eq:eom}
    \bar{m}_1 \frac{dv_1^{\mu}}{d\tau_1} = -\eta^{\mu \nu} \Bigg(\sum_{n=1}^{\infty} \frac{\partial \mathcal{L}_n}{\partial x_1^{\nu}(\tau_1)} - \frac{d}{d\tau_1} \Big(\frac{\partial \mathcal{L}_n}{\partial v_1^{\nu}}\Big)\Bigg)~,
\end{align}
from the tree-level action is,
\begin{align}
\frac{dv_1^{\mu}}{d\tau_1}=&\frac{ \bar{m}_2}{2\widetilde{M}_{\rm pl}^2}\int d\tau_2\left(2(v_2(\tau_2)\cdot v_1(\tau_1))^2-\frac{2}{d-2}v_1^2(\tau_1)v_2^2(\tau_2)\right)\int_k  \frac{ik^{\mu}}{k^2}e^{i k\cdot(x_1(\tau_1)-x_2(\tau_2))}
\nonumber \\&
- \frac{\bar{m}_2}{\widetilde{M}_{\rm pl}^2}\int d\tau_2 \left(2(v_2(\tau_2)\cdot v_1(\tau_1))v_2^{\mu}(\tau_2)-\frac{2}{d-2}v_1^{\mu}(\tau_1)v_2^2(\tau_2)\right)\int_k\frac{ik\cdot v_1(\tau_1)}{k^2}e^{i k\cdot(x_1(\tau_1)-x_2(\tau_2))}
\nonumber \\ &
- \frac{\bar{m}_2}{\widetilde{M}_{\rm pl}^2}\int d\tau_2 \left(2(v_2(\tau_2)\cdot \frac{dv_1(\tau_1)}{d\tau_1})v_2^{\mu}(\tau_2)-\frac{2}{d-2}\frac{dv_1^{\mu}(\tau_1)}{d\tau_1}v_2^2(\tau_2)\right)\int_k\frac{1}{k^2}e^{i k\cdot(x_1(\tau_1)-x_2(\tau_2))}
\nonumber \\
&+\frac{ \bar{m}_2}{4\widetilde{M}_{\rm pl}^2}\int d\tau_2  \left(\tilde{d}_1^{(1)}v_1^2(\tau_1)+\tilde{f}_1^{(1)}\right)\left(\tilde{d}_1^{(2)}v_2^2(\tau_2)+\tilde{f}_1^{(2)}\right)\int_k \frac{ik^{\mu}}{k^2}e^{i k\cdot(x_1(\tau_1)-x_2(\tau_2))}
\nonumber\\
&-\frac{ \bar{m}_2}{2\widetilde{M}_{\rm pl}^2}\int d\tau_2\  \tilde{d}_1^{(1)}v_1^{\mu}(\tau_1)\left(\tilde{d}_1^{(2)}v_2^2(\tau_2)+\tilde{f}_1^{(2)}\right)\int_k \frac{ik\cdot v_1(\tau_1)}{k^2}e^{i k\cdot(x_1(\tau_1)-x_2(\tau_2))}
\nonumber \\
&-\frac{ \bar{m}_2}{2\widetilde{M}_{\rm pl}^2}\int d\tau_2\ \tilde{d}_1^{(1)} \frac{dv_1^{\mu}(\tau_1)}{d\tau_1}\left(\tilde{d}_1^{(2)}v_2^2(\tau_2)+\tilde{f}_1^{(2)}\right)\int_k \frac{1}{k^2}e^{i k\cdot(x_1(\tau_1)-x_2(\tau_2))}~.
\label{eq:eom1pm}
\end{align}
These EOMs will be used below to compute the corrections to the trajectories, which in turn will be important for computing the next-to-leading-order corrections $\mathcal{O}(G)^{(n>1)}$ to the impulse and hence, the scattering angle.

\subsection{Next-to-leading-order}
The effective Lagrangian at the next-to-leading-order (NLO), $\mathcal{O}(G)^2$ (or `one-loop'), for the remaining diagrams shown in Fig.~\ref{diag:Feynman2PM} is,

\begin{align}
    \mathcal{L}_2 =&\frac{\bar{m}_1\bar{m}_2^2}{2\widetilde{M}_{\rm pl}^3}v_1^{\alpha}(\tau_1)v_1^{\beta}(\tau_1)\int d\tau_2\int d\bar{\tau}_2v_2^{\mu}(\tau_2)v_2^{\nu}(\tau_2)v_2^{\lambda}(\bar{\tau}_2)v_2^{\rho}(\bar{\tau}_2)\mathcal{P}_{\alpha \beta \bar{\alpha}\bar{\beta}}(k_1)\mathcal{P}_{\mu \nu \bar{\mu}\bar{\nu}}(k_2)\mathcal{P}_{\lambda \rho \bar{\lambda}\bar{\rho}}(k_3)\nonumber\\ &\times\int_{k_{1,2,3}}e^{ik_1x_1(\tau_1)}e^{ik_2x_2(\tau_2)}e^{ik_3x_2(\bar{\tau}_2)}\frac{V_{hhh}^{\bar{\alpha}\bar{\beta}\bar{\mu}\bar{\nu}\bar{\lambda}\bar{\rho}}(k_1,k_2,k_3)}{k_1^2k_2^2k_3^3}\delta^4(k_1+k_2+k_3)
    \nonumber\\
    &+ \frac{\bar{m}_1\bar{m}_2^2}{16\widetilde{M}_{\rm pl}^3}\int d\tau_2 \int d\bar{\tau}_2 \left(\tilde{d}_1^{(1)}v_1^2(\tau_1)+\tilde{f}_1^{(1)}\right)\left(\tilde{d}_1^{(2)}v_2^2(\tau_2)+\tilde{f}_1^{(2)}\right)\left(\tilde{d}_1^{(2)}v_2^2(\bar{\tau}_2)+\tilde{f}_1^{(2)}\right) 
    \nonumber \\
    &\times\int_{k_{1,2,3}}\frac{V_{\psi \psi \psi}(k_1,k_2,k_3)}{k_1^2 k_2^2 k_3^2} e^{ik_1x_1(\tau_1)}e^{ik_2x_2(\tau_2)}e^{ik_3x_2(\bar{\tau}_2)}\delta^4(k_1+k_2+k_3)
    \nonumber \\
    &+\frac{\bar{m}_1\bar{m}_2^2}{8\widetilde{M}_{\rm pl}^3}v_1^{\mu}(\tau_1)v_1^{\nu}(\tau_1)\int d\tau_2 \int d\bar{\tau}_2\left(\tilde{d}_1^{(2)}v_2^2(\tau_2)+\tilde{f}_1^{(2)}\right)\left(\tilde{d}_1^{(2)}v_2^2(\bar{\tau}_2)+\tilde{f}_1^{(2)}\right)
    \nonumber \\
    &\times\int_{k_{1,2,3}}\frac{\mathcal{P}_{\mu \nu \alpha \beta}(k_1)}{k_1^2k_2^2k_3^2}V^{\alpha \beta}_{\psi \psi h}(k_2,k_3)e^{ik_1x_1(\tau_1)}e^{ik_2x_2(\tau_2)}e^{ik_3x_2(\bar{\tau}_2)}\delta^4(k_1+k_2+k_3)
    \nonumber \\
    &+\frac{\bar{m}_1\bar{m}_2^2}{4\widetilde{M}_{\rm pl}^3}\left(\tilde{d}_1^{(1)}v_1^2(\tau_1)+\tilde{f}_1^{(1)}\right)\int d\tau_2 \int d\bar{\tau}_2\left(\tilde{d}_1^{(2)}v_2^2(\tau_2)+\tilde{f}_1^{(2)}\right)v_2^{\mu}(\bar{\tau}_2)v_2^{\nu}(\bar{\tau}_2) 
    \nonumber \\
    &\times\int_{k_{1,2,3}}\frac{\mathcal{P}_{\mu \nu \alpha \beta}(k_3)}{k_1^2k_2^2k_3^2}V^{\alpha \beta}_{\psi\psi h}(k_1,k_2)e^{ik_1x_1(\tau_1)}e^{ik_2x_2(\tau_2)}e^{ik_3x_2(\bar{\tau}_2)}\delta^4(k_1+k_2+k_3)
    \nonumber \\
    &-\frac{\bar{m}_1\bar{m}_2^2}{16\widetilde{M}_{\rm pl}^4}\left(\tilde{d}_2^{(1)}v_1^2(\tau_1)+\tilde{f}_2^{(1)}\right)\int d\tau_2 \int d\bar{\tau}_2 \left(\tilde{d}_1^{(2)}v_2^2(\tau_2)+\tilde{f}_1^{(2)}\right)\left(\tilde{d}_1^{(2)}v_2(\bar{\tau}_2)^2+\tilde{f}_1^{(2)}\right)\nonumber \\
    &\times\int_{k_{1,2}}\frac{1}{k_1^2 k_2^2}e^{i(k_1+k_2)x_1(\tau_1)}e^{-ik_1x_2(\tau_2)}e^{-ik_2x_2(\bar{\tau}_2)}
    \nonumber\\
    & -\frac{\bar{m}_1\bar{m_2}^2}{2 \widetilde{M}_{\rm pl}^4}\tilde{d}_1^{(1)}v_1^{\mu}(\tau_1)v_1^{\nu}(\tau_1)\int d\tau_2 \int d\bar{\tau}_2 v_2^{\alpha}(\tau_2)v_2^{\beta}(\tau_2)\left(\tilde{d}_1^{(2)}v_2(\bar{\tau}_2)^2+\tilde{f}_1^{(2)}\right)
    \nonumber\\
    &\times \int_{k_{1,2}}\frac{\mathcal{P}_{\mu \nu \alpha \beta}(k_1)}{k_1^2k_2^2}e^{i(k_1+k_2)x_1(\tau_1)}e^{-ik_1x_2(\tau_2)}e^{-ik_2x_2(\bar{\tau}_2)}
    \nonumber\\
    &+ 1\leftrightarrow 2~.
    \label{eq:L2PM}
\end{align}
The bulk vertices $V_{\psi \psi\psi}(k_1,k_2,k_3)$, $V^{\alpha \beta}_{\psi\psi h}(k_1,k_2)$ and $V_{hhh}(k_1,k_2,k_3)$ are given in Table~\ref{tab:Feynman}. Also, the last two contributions in the above equation are due to the worldline non-linearities, i.e. the scalar-scalar and scalar-graviton worldline vertice as given in Table~\ref{tab:Feynman}.

\subsection{Next-to-next-to-leading-order}
The effective Lagrangian at the next-to-next-to-leading-order (NNLO), $\mathcal{O}(G)^3$ (or `two-loop'), is computed from the Feynman diagrams listed in Appendix~\ref{app:Feynman3PM}. The explicit expression for the Lagrangian $\mathcal{L}_3$ can be derived using the Feynman rules given in Table~\ref{tab:Feynman}. At first sight, there are many diagrams at 3PM order, but the orthogonality condition and the vanishing of scaleless integrals provide strong kinematic and dimensional selection rules which drastically reduce the number of contributions when the diagrams are evaluated on the undeflected trajectories. 

\subsection{Trajectories}
To obtain the trajectories, we now integrate the above EOMs using the weak-field PM expansion of the velocity and trajectory around the straight line motion, namely,
\begin{align}
    v_a^{\mu} (\tau_a) &= u_a^{\mu} + \sum_n \delta^{(n)}v_a^{\mu} (\tau_a)~, \nn \\
   x_a^{\mu} (\tau_a) &= b_a^{\mu} + u_a^{\mu} \tau_a + \sum_n \delta^{(n)}x_a^{\mu} (\tau_a)~, 
\end{align}
where $u_a^{\mu}$ is the incoming velocity at infinity and $b_a^{\mu}$ is related to the initial position of each particle. Since for the scattering processes, the particles are assumed at asymptotically large separations, the condition in Eq.~\eqref{eq:einbein} implies that $u_a^2 = 1$ with $u_1\cdot u_2 = \gamma $. Moreover, the initial position vector is such that it is orthogonal to the four-velocity $u_a$, and thus the parameter $b^{\mu} \equiv b^{\mu}_1 - b^{\mu}_2$ will be associated with the impact parameter of the scattering configuration in the centre-of-mass (COM) frame.

Since, we compute the impulse up to 3PM order, we need the LO and NLO corrections to the unperturbed solution.
\subsubsection{Leading Order}
Using the 1PM EOMs given in Eq.~\eqref{eq:eom1pm}, the first correction to the undeflected velocity of particle 1 in $d$-dimensions is,
\begin{align}
\label{eq:vel1PM}
    \delta^{(1)}v_1^{\mu}(\tau_1) &= \frac{\bar{m}_2}{\widetilde{M}_{\rm pl}^2}\left[\Bigg(\frac{2\gamma^2-\frac{2}{d-2}}{2}+\frac{\left(\tilde{d}_1^{(1)}+\tilde{f}_1^{(1)}\right)\left(\tilde{d}_1^{(2)}+\tilde{f}_1^{(2)}\right)}{4}\Bigg)\eta^{\mu \alpha}- \Bigg(2\gamma u_2^{\mu} - \frac{2}{d-2}u_1^{\mu}\Bigg)u_1^{\alpha}
    \right.\nn\\
    & \left.
    -\frac{\tilde{d}_1^{(1)}\left(\tilde{d}_1^{(2)}+\tilde{f}_1^{(2)}\right)}{2}u_1^{\mu} u_1^{\alpha}\right]\int_{-\infty}^{\tau_1}d\tilde{\tau}_1\int_k\frac{ik_{\alpha}}{k^2}\hat{\delta}(u_2\cdot k)e^{ik\cdot b}e^{i(k\cdot u_1-i\epsilon)\tilde{\tau}_1}~,
    \nonumber \\
    & = \frac{\bar{m}_2}{\widetilde{M}_{\rm pl}^2}\left[\Bigg(\frac{2\gamma^2-\frac{2}{d-2}}{2}+\frac{\big(\tilde{d}_1^{(1)}+\tilde{f}_1^{(1)}\big)\big(\tilde{d}_1^{(2)}+\tilde{f}_1^{(2)}\big)}{4}\Bigg)\eta^{\mu \alpha}-2\gamma u_2^{\mu} u_1^{\alpha} 
    \right. \nn\\
    & \left.- \Bigg(\frac{\tilde{d}_1^{(1)}\left(\tilde{d}_1^{(2)}+\tilde{f}_1^{(2)}\right)}{2}-\frac{2}{d-2}\Bigg)u_1^{\mu} u_1^{\alpha}\right]\int_k\frac{k_{\alpha}}{k^2}\hat{\delta}(u_2\cdot k)e^{ik\cdot b}\frac{e^{i(k\cdot u_1-i\epsilon){\tau}_1}}{(k\cdot u_1 -i\epsilon)}~,
\end{align}
where $\hat{\delta}(x) = 2 \pi{\delta}(x)$, we hereafter use this notation to absorb the factors of $2\pi.$
The correction to the undeflected position of the particle 1 is,
\begin{align}
\label{eq:pos1PM}
    \delta^{(1)}x_1^{\mu}(\tau_1) =& -i\frac{\bar{m}_2}{\widetilde{M}_{\rm pl}^2}\left[\Bigg(\frac{2\gamma^2-\frac{2}{d-2}}{2}+\frac{\big(\tilde{d}_1^{(1)}+\tilde{f}_1^{(1)}\big)\big(\tilde{d}_1^{(2)}+\tilde{f}_1^{(2)}\big)}{4}\Bigg)\eta^{\mu \alpha}-2\gamma u_2^{\mu} u_1^{\alpha} 
    \right. \nn\\
    & \left.- \Bigg(\frac{\tilde{d}_1^{(1)}\left(\tilde{d}_1^{(2)}+\tilde{f}_1^{(2)}\right)}{2}-\frac{2}{d-2}\Bigg)u_1^{\mu} u_1^{\alpha}\right]\int_k\frac{k_{\alpha}}{k^2}\hat{\delta}(u_2\cdot k)e^{ik\cdot b}\frac{e^{i(k\cdot u_1-i\epsilon){\tau}_1}}{(k\cdot u_1 -i\epsilon)^2}~.
\end{align}
We have added $-i \epsilon$ to ensure the convergence of the integral as $\tilde{\tau}_1 \rightarrow -\infty$. The deflection of the second particle is obtained by following the same procedure, and due to the symmetry of the setup, is given simply by the relabelling of $1\leftrightarrow 2$ and $k \rightarrow -k$ in the corrections to particle 1. To ensure the convergence of the corresponding integrals for particle 2, the factor $(k\cdot u_1 -i\epsilon)$ turns into $(k\cdot u_2 +i\epsilon)$. 
\subsubsection{Next-to-leading order}
The NLO corrections to the trajectories receives two types of contributions: i) from the EOMs corresponding to the 2PM action \eqref{eq:L2PM} computed on the undeflected trajectories, $\delta^{(2)}_{\mathcal{L}_2}$, and; ii) from the iterations of the 1PM EOMs given in  Eq.~\eqref{eq:eom1pm} with the above computed 1PM corrections to trajectory, $\delta^{(2)}_{\mathcal{L}_1}$. The total 2PM correction to the trajectory is, therefore, $\delta^{(2)} = \delta^{(2)}_{\mathcal{L}_1}+ \delta^{(2)}_{\mathcal{L}_2}$.

First, we discuss the contribution to $\delta^{(2)}_{\mathcal{L}_2}$. The EOMs associated with the 2PM effective action originate from one‑loop diagrams in which two bulk fields can attach to the same worldline at two distinct proper times. Consequently, the 2PM correction to the trajectory of particle 1 must include all insertions of the corresponding effective operator on its worldline, i.e. both contributions from the two distinct proper times associated with the same worldline leg of particle 1. The correction to the unperturbed velocity of particle 1 due to $\mathcal{L}_2$ is then,
\begin{align}
\label{eq:EOM2PM}
          \delta^{(2)}_{\mathcal{L}_2} v_1^{\mu}(\tau_1) = & -\frac{1}{\bar{m}_1} \eta^{\mu \nu} \Bigg( \frac{\partial \mathcal{L}_2}{\partial x_1^{\nu}(\tau_1)} - \frac{d}{d\tau_1} \Big(\frac{\partial \mathcal{L}_2}{\partial v_1^{\nu}(\tau_1)}\Big) \Bigg) \nn\\
          &-\frac{1}{\bar{m}_1} \eta^{\mu \nu} \Bigg( \frac{\partial \mathcal{L}_2}{\partial x_1^{\nu}(\tilde{\tau}_1)} - \frac{d}{d\tilde{\tau}_1} \Big(\frac{\partial \mathcal{L}_2}{\partial v_1^{\nu}(\tilde{\tau}_1)}\Big)\Bigg)~.
\end{align}
Here,
$\tilde{\tau}_1$ and ${\tau}_1$ 	
label the two insertions of the same worldline operator on particle 1.  
For the purely gravitational sector, the integrand is symmetric under ${\tau}_1\rightarrow \tilde{\tau}_1$, so the two contributions are equal and may be combined into an overall factor of two. This correction, $\delta^{(2)}_{\mathcal{L}_2}$, can be obtained straightforwardly by inserting the 2PM Lagrangian~\eqref{eq:L2PM} into the EOMs~\eqref{eq:EOM2PM}, evaluated on undeflected (straight-line) trajectories.

The second contribution, $\delta^{(2)}_{\mathcal{L}_1}$ can be computed by inserting the 1PM trajectories \eqref{eq:vel1PM}-\eqref{eq:pos1PM} into the 1PM EOMs \eqref{eq:eom1pm} and keeping terms up to $\mathcal{O}(G^2)$ only. For simplicity, we omit the explicit computations here.

\section{\label{sec:results} Results}
In this section, we present our results for the gauge invariant conservative scattering angle for hyperbolic encounters up to 3PM order. We explicitly display the impulse for particle 1
\begin{equation}
\label{eq:imp}
\Delta p_1^{\mu} = -\eta_{\mu \nu}\int d\tau_1 \frac{\partial \mathcal{L}_1}{\partial x_1^{\nu}}~;
\end{equation}
while the result for particle 2 follows by exchanging the particle labels. To simplify the integrals entering in the impulse computation, we will work in the rest frame where one of the incoming particle is at rest, and align the initial velocity of the companion with the $x$-axis, so that the perpendicular direction of the momentum $k_{\perp}$ is in the $yz$-plane. Once the impulse is computed, we will then directly read off the scattering angle by going to the COM frame using
\begin{align}
    2\sin \left(\frac{\chi}{2}\right) = \chi -\frac{1}{24} \chi^3 + \frac{1}{1920} \chi^5 + \mathcal{O}(\chi^7) = \frac{\Delta p_{1 cm}}{p_{\infty}^{cm}} = \frac{\sqrt{-\Delta p^2}}{p_{\infty}^{cm}}~,
    \label{eq:scat}
\end{align}
where $\chi$ is the scattering angle
and $p_{\infty}^{cm}$ is the COM relative momentum at infinity given by, 
\begin{align}
    p_{\infty}^{cm} = \mu \frac{\sqrt{\gamma^2-1}}{\Gamma}~,
\end{align}
with $\Gamma \equiv E/M = \sqrt{1+2\nu (\gamma-1)}$, $\mu = \bar{m}_1\bar{m}_2/(\bar{m}_1+\bar{m}_2)$ is the reduced mass, and $\nu = \mu/(\bar{m}_1+\bar{m}_2)$ is the symmetric mass-ratio. The PM coefficients of the scattering angle $\chi$ as an expansion in impact parameter $b$ are defined via,
\begin{align}
    \frac{\chi}{2} = \sum_{n} \chi_b^{(n)}\left(\frac{G M}{\phi_0b}\right)^n~.
\end{align}
\subsection{\label{sec:1PM} Leading order: the 1PM scattering angle}

The LO impulse, evaluated on the undeflected trajectory and obtained by inserting the LO Lagrangian~\eqref{1PML} into Eq.~\eqref{eq:imp} is,
\begin{align}
\Delta^{(1)}_{\mathcal{L}_1} p_1^{\mu} &= \frac{\bar{m}_1\bar{m}_2}{2\widetilde{M}_{\rm pl}^2}(2\gamma^2-1)\int_kik^{\mu} \frac{\hat{\delta}(k\cdot u_1)\hat{\delta}(k\cdot u_2)}{k^2}e^{ik\cdot b}\nonumber \\ & +\frac{\bar{m}_1\bar{m}_2}{4\widetilde{M}_{\rm pl}^2}\left(\tilde{d}_1^{(1)}+\tilde{f}_1^{(1)}\right)\left(\tilde{d}_1^{(2)}+\tilde{f}_1^{(2)}\right)\int_kik^{\mu}\frac{\hat{\delta}(k\cdot u_1)\hat{\delta}(k\cdot u_2)}{k^2}e^{ik\cdot b}~.
\end{align}
By going to the rest frame of particle 1, i.e. $u_1 = (1, 0, 0, 0)$ and $u_2 = (\gamma, \gamma \beta, 0, 0)$ with $\beta = \sqrt{\gamma^2-1}/\gamma$, we obtain,
\begin{align}
\Delta^{(1)}_{\mathcal{L}_1}  p_1^{\mu} &=- \frac{\bar{m}_1\bar{m}_2}{2\widetilde{M}_{\rm pl}^2}\frac{(2\gamma^2-1)}{\sqrt{\gamma^2-1}}\frac{b^{\mu}}{2\pi |b^2|}\nonumber \\ & -\frac{\bar{m}_1\bar{m}_2}{4\widetilde{M}_{\rm pl}^2}\left(\tilde{d}_1^{(1)}+\tilde{f}_1^{(1)}\right)\left(\tilde{d}_1^{(2)}+\tilde{f}_1^{(2)}\right)\frac{1}{\sqrt{\gamma^2-1}}\frac{b^{\mu}}{2\pi |b^2|}\\
& = -\frac{2\bar{m}_1\bar{m}_2 G}{\phi_0}\frac{1}{\sqrt{\gamma^2-1}}\left(2\gamma^2-1+\frac{\left(\tilde{d}_1^{(1)}+\tilde{f}_1^{(1)}\right)\left(\tilde{d}_1^{(2)}+\tilde{f}_1^{(2)}\right)}{2}\right)\frac{b^{\mu}}{|b^2|}~.
\label{eq:imp1PM}
\end{align}
Inserting the impulse in~\eqref{eq:scat}, the LO scattering angle is,
\begin{align}
    \frac{\chi_b^{(1)}}{\Gamma} =  \frac{1}{(\gamma^2-1)}\left(2\gamma^2-1+\frac{\left(\tilde{d}_1^{(1)}+\tilde{f}_1^{(1)}\right)\left(\tilde{d}_1^{(2)}+\tilde{f}_1^{(2)}\right)}{2}\right)~.
\end{align}

\subsection{\label{sec:2PM} Next-to-leading-order: 2PM scattering angle}

At the 2PM order, the impulse and thus the scattering angle have two distinct contributions. The first one comes from the 2PM action computed at the undeflected trajectory, $\Delta^{(2)}_{\mathcal{L}_2}p^{\mu}_1$, and the second one originates from the 1PM action computed at the perturbed (LO) trajectory, $\Delta^{(2)}_{\mathcal{L}_1}p^{\mu}_1$. 

The contributions to the impulse from the LO corrections in position and velocity is,
\begin{align}
    \Delta^{(2)}_{\mathcal{L}_1}p^{\mu}_1 =&\frac{\bar{m}_1\bar{m}_2}{\widetilde{M}_{\rm pl}^2}\int_{-\infty}^{\infty}d\tau_1 d\tau_2 \int_k\frac{ik^{\mu}}{k^2}
    \left\{\Bigg(2\gamma u_2+u_1\Bigg(\frac{\tilde{d}_1^{(1)}\Big(\tilde{d}_1^{(2)}+\tilde{f}_1^{(2)}\Big)}{2}-1\Bigg)\Bigg)\cdot \delta^{(1)}v_1(\tau_1)
    \right.\nn\\
    &\left.+\left(\frac{2\gamma^2-1}{2}+\frac{\left(\tilde{d}_1^{(1)}+\tilde{f}_1^{(1)}\right)\left(\tilde{d}_1^{(2)}+\tilde{f}_1^{(2)}\right)}{4}\right)(ik)\cdot \left(\delta^{(1)}x_1(\tau_1)-\delta^{(1)}x_2(\tau_2)\right)\right.\nonumber\\
    &\left.+\Bigg(2\gamma u_1+u_2\Bigg(\frac{\Big(\tilde{d}_1^{(1)}+\tilde{f}_1^{(1)}\Big)\tilde{d}_1^{(2)}}{2}-1\Bigg)\Bigg)\cdot \delta^{(1)}v_2(\tau_2)\right\}e^{ik\cdot b+ik\cdot (u_1\tau_1-u_2\tau_2)}~.
\end{align}
To simplify our analysis, we only consider the deflected motion of particle 1, and consider the mirrory images only when combining the final result. Using Eqs.~\eqref{eq:vel1PM}-\eqref{eq:pos1PM} in the above equation, we obtain,
\begin{align}
    \Delta^{(2)}_{\mathcal{L}_1}p^{\mu}_1 = & i\frac{\bar{m}_1\bar{m}_2^2}{\widetilde{M}_{\rm pl}^4}\int_{\ell,k}\left\{\left(2\gamma^2-1+\frac{\left(\tilde{d}_1^{(1)}+\tilde{f}_1^{(1)}\right)\left(\tilde{d}_1^{(2)}+\tilde{f}_1^{(2)}\right)}{2}\right)^2\frac{\ell^2}{8}\right.\nonumber\\
    &\left.-2\gamma^2\left(1+\frac{\tilde{d}_1^{(1)}\left(\tilde{d}_1^{(2)}+\tilde{f}_1^{(2)}\right)}{2}\right)(k\cdot u_1)^2-\frac{\tilde{f}_1^{(1)}\left(\tilde{d}_1^{(2)}+\tilde{f}_1^{(2)}\right)}{2}\left(1-\frac{\tilde{d}_1^{(1)}\left(\tilde{d}_1^{(2)}+\tilde{f}_1^{(2)}\right)}{2}\right)(k\cdot u_1)^2\right\}\nonumber\\
    &\times \frac{(\ell^{\mu}-k^{\mu})\hat{\delta}(\ell\cdot u_1)\hat{\delta}(\ell\cdot u_2)\hat{\delta}(k\cdot u_2)}{k^2(\ell-k)^2(k\cdot u_1-i\epsilon)^2}e^{i\ell\cdot b}\,.
\end{align}

Now, the second contribution $\Delta^{(2)}_{\mathcal{L}_2}p^{\mu}_1$ coming from the 2PM action computed at the undeflected trajectory is,
\begin{align}
    \Delta_{\mathcal{L}_2}^{(2)}p_1^{\mu} = &\frac{i\bar{m}_1\bar{m}_2^2(\gamma^2+3)}{16\widetilde{M}_{\rm pl}^4}\int_{\ell, k}\frac{\ell^{\mu}\delta(\ell \cdot u_1)\delta(\ell \cdot u_2)\delta(k\cdot u_2)}{k^2(\ell-k)^2}e^{i\ell\cdot b} 
    \nonumber \\
    & -\frac{i\bar{m}_1\bar{m}_2^2}{32\widetilde{M}_{\rm pl}^4}c_1\left(\tilde{d}_1^{(1)}+\tilde{f}_1^{(1)}\right)\left(\tilde{d}_1^{(2)}+\tilde{f}_1^{(2)}\right)^2\int_{\ell, k}\frac{\ell^{\mu}\delta(\ell \cdot u_1)\delta(\ell \cdot u_2)\delta(k\cdot u_2)}{k^2(\ell-k)^2}e^{i\ell\cdot b}
    \nonumber\\
    &+\frac{i\bar{m}_1\bar{m}_2^2(\gamma^2-1)}{32\widetilde{M}_{\rm pl}^4}\left(\tilde{d}_1^{(2)}+\tilde{f}_1^{(2)}\right)^2\int_{\ell, k}\frac{\ell^{\mu}\delta(\ell \cdot u_1)\delta(\ell \cdot u_2)\delta(k\cdot u_2)}{k^2(\ell-k)^2}e^{i\ell\cdot b}
    \nonumber\\
    &+\frac{i\bar{m}_1\bar{m}_2^2}{16\widetilde{M}_{\rm pl}^4}\left(\tilde{d}_2^{(1)}+\tilde{f}_2^{(1)}\right)\left(\tilde{d}_1^{(2)}+\tilde{f}_1^{(2)}\right)^2\int_{\ell, k}\frac{\ell^{\mu}\delta(\ell \cdot u_1)\delta(\ell \cdot u_2)\delta(k \cdot u_2)}{k^2(\ell-k)^2}e^{i\ell\cdot b}
    \nonumber \\
    &+\frac{i\bar{m}_1\bar{m}_2^2}{4\widetilde{M}_{\rm pl}^4}(2\gamma^2-1)\tilde{d}_1^{(1)}\left(\tilde{d}_1^{(2)}+\tilde{f}_1^{(2)}\right)\int_{\ell,k}\frac{\ell^{\mu}\delta(\ell \cdot u_1)\delta(\ell \cdot u_2)\delta(k \cdot u_2)}{k^2(\ell-k)^2}e^{i\ell\cdot b}~.
\end{align}
Following the same procedure as in~\cite{Kalin:2020fhe}, we split the above two contributions into more ``suggestive" contributions as, 
\begin{align}
    \Delta^{(2)}p_1^{\mu} &= \Delta_{\mathcal{L}_1}^{(2)}p_1^{\mu} +\Delta_{\mathcal{L}_2}^{(2)}p_1^{\mu} \nonumber \\
    &= \Delta_{b}^{(2)}p_1^{\mu} +\Delta_{u}^{(2)}p_1^{\mu}~,     
\end{align}
where $\Delta_{b}^{(2)}p_1^{\mu}$ is the piece proportional to impact parameter $b^{\mu}$ and $\Delta_{u}^{(2)}p_1^{\mu}$ is the remaining part, proportional to the combination of four-velocities $u_a^{\mu}$ after performing the integrations. The $\Delta_{u}^{(2)}p_1^{\mu}$ piece arises purely from the iterations of the 1PM impulse with the 1PM trajectories, whereas $\Delta_{b}^{(2)}p_1^{\mu}$ arises from both contributions. This decomposition can also be verified using the momentum conservation and the on-shell condition,
\begin{align}
\label{eq:iterR}
    (p_a+\Delta p_a)^2 = p_a^2 \implies 2p_a\cdot \Delta p_a = - \Delta p_a^2~,
\end{align}
together with the facts that $\Delta^{(1)}_{\mathcal{L}_1}p^{\mu}\propto b^{\mu}$ at 1PM order (see Eq.~\eqref{eq:imp1PM}), and
$b\cdot u_a = 0$. From the above relation, we can iteratively solve for the components along the velocities, thereby showing that the term proportional to the velocities is generated purely by iteration effects. 

The two pieces at 2PM order in impulse are,
\begin{align}
\label{eq:PM2-del}
    \Delta_{b}^{(2)}p_1^{\mu} &=\frac{\bar{m}_1\bar{m}_2^2}{16\widetilde{M}_{\rm pl}^4}\Bigg\{-3(5\gamma^2-1)-4\left(\tilde{d}_1^{(1)}+\tilde{f}_1^{(1)}\right)\left(\tilde{d}_1^{(2)}+\tilde{f}_1^{(2)}\right)+\frac{\left(\tilde{d}_1^{(2)}+\tilde{f}_1^{(2)}\right)^2}{2}\left[\gamma^2-1
    \right.\nonumber\\
    &\left.+2\left(\tilde{d}_2^{(1)}+\tilde{f}_2^{(1)}\right)+4\tilde{d}_1^{(1)}\tilde{f}_1^{(1)}-c_1 \left(\tilde{d}_1^{(1)}+\tilde{f}_1^{(1)}\right)\right]\Bigg\}\frac{\partial}{\partial b_{\mu}}\int_{\ell,k}\frac{\delta(\ell \cdot u_1)\delta(\ell \cdot u_2)\delta(k \cdot u_2)}{k^2(\ell-k)^2}e^{i\ell\cdot b}~,
\end{align}
and 
\begin{align}
\label{eq:PM2-u}
    \Delta_{u}^{(2)}p_1^{\mu} &=i\frac{\bar{m}_1\bar{m}_2^2}{8\widetilde{M}_{\rm pl}^4}\Bigg(2\gamma^2-1+\frac{\Big(\tilde{d}_1^{(1)}+\tilde{f}_1^{(1)}\Big)\Big(\tilde{d}_1^{(2)}+\tilde{f}_1^{(2)}\Big)}{2}\Bigg)^2\int_{\ell,k}\frac{(\ell^{\mu}-k^{\mu})\ell^2\hat{\delta}(\ell\cdot u_1)\hat{\delta}(\ell\cdot u_2)\hat{\delta}(k\cdot u_2)}{k^2(\ell-k)^2(k\cdot u_1-i\epsilon)^2}e^{i\ell\cdot b}~.
\end{align}
 
To solve the integrals in the impulse, we go to the rest frame of the particle 2 (instead of particle 1) and obtain the result,
\begin{align}
\label{eq:impb}
    \Delta_{b}^{(2)}p_1^{\mu} &=-\frac{\pi\bar{m}_1\bar{m}_2^2}{4\sqrt{\gamma^2-1}}\frac{G^2b^{\mu}}{\phi_0^2|b^2|^{3/2}}\biggl\{3(5\gamma^2-1)+4\left(\tilde{d}_1^{(1)}+\tilde{f}_1^{(1)}\right)\Big(\tilde{d}_1^{(2)}+\tilde{f}_1^{(2)}\Big) \nonumber\\
    & -\frac{\Big(\tilde{d}_1^{(2)}+\tilde{f}_1^{(2)}\Big)^2}{2}\left[2\left(\tilde{d}_2^{(1)}+\tilde{f}_2^{(1)}\right)-c_1 \alpha_1+4\tilde{d}_1^{(1)}\tilde{f}_1^{(1)}+\gamma^2-1\right]\biggr\}+ 1\leftrightarrow 2~.
\end{align}
Solving the second integral~\eqref{eq:PM2-u} as in~\cite{Kalin:2020mvi}, we obtain 
\begin{align}
\label{eq:impu}
    \Delta_{u}^{(2)}p_1^{\mu} &=2\frac{\bar{m}_1\bar{m}_2^2}{(\gamma^2-1)^2}\left(2\gamma^2-1+\frac{\Big(\tilde{d}_1^{(1)}+\tilde{f}_1^{(1)}\Big)\Big(\tilde{d}_1^{(2)}+\tilde{f}_1^{(2)}\Big)}{2}~\right)^2\frac{G^2}{\phi_0^2|b^2|}(\gamma u_2^{\mu}-u_1^{\mu})- 1\leftrightarrow 2~,
\end{align}
which agrees with the result obtained from iteratively solving the relation \eqref{eq:iterR}. The antisymmetrization in $1\leftrightarrow2$ in the last contribution is because the deflection of particle 2 shifts the pole from $i \epsilon$ to $-i \epsilon$, which shifts the integral from the upper complex half-plane to the lower complex half-plane, resulting in an overall negative sign. 

Finally, using the impulse in Eq.~\eqref{eq:scat}, the NLO scattering angle is,
\begin{align}
    \frac{\chi_b^{(2)}}{\Gamma} = &\frac{\pi}{8 (\gamma^2-1)}\Bigg(3(5\gamma^2-1)+4\left(\tilde{d}_1^{(1)}+\tilde{f}_1^{(1)}\right)\Big(\tilde{d}_1^{(2)}+\tilde{f}_1^{(2)}\Big)
    \nn\\
    &-\frac{\bar{m}_2\Big(\tilde{d}_1^{(2)}+\tilde{f}_1^{(2)}\Big)^2}{2M}\left[2\left(\tilde{d}_2^{(1)}+\tilde{f}_2^{(1)}\right)-c_1 \left(\tilde{d}_1^{(1)}+\tilde{f}_1^{(1)}\right)+4\tilde{d}_1^{(1)}\tilde{f}_1^{(1)}+\gamma^2-1\right]
    \nn\\
    &-\frac{\bar{m}_1\Big(\tilde{d}_1^{(1)}+\tilde{f}_1^{(1)}\Big)^2}{2M}\left[2\left(\tilde{d}_2^{(2)}+\tilde{f}_2^{(2)}\right)-c_1 \left(\tilde{d}_1^{(2)}+\tilde{f}_1^{(2)}\right)+4\tilde{d}_1^{(2)}\tilde{f}_1^{(2)}+\gamma^2-1\right]\Bigg)~.
\end{align}

\subsection{\label{sec:3PM} Next-to-next-to-leading-order: 3PM scattering angle}

At the 3PM order, similar to the 2PM order, the total impulse and thus the scattering angle are obtained from three distinct contributions; the first one from the 3PM action computed at the undeflected trajectory and the two other ones from the iterations of the 2PM and 1PM impulses with LO and NLO trajectories, respectively. As mentioned previously, the velocity components of the impulse can be obtained by iteratively solving the relation \eqref{eq:iterR} using the impulse at 1PM and 2PM order, thereby allowing us to restrict the derivation of the impulse in the perpendicular plane (i.e. proportional to impact parameter). Emploing the usual techniques of decomposition to a basis of Master Integrals, we end up with the same basis as the one in the 3PM computation in the spinless sector of pure GR, as reported in~\cite{Jakobsen:2022fcj}.  The $b$-direction component of the impulse at 3PM is,
\begin{align}
        \Delta_{b}^{(3)}p_1^{\mu} = &\frac{G^3 b^{\mu}}{\phi_0^3|b^2|^2} \Bigg[\frac{8m_1^2 m_2^2 \sinh^{-1}\sqrt\frac{\gamma-1}{2}}{(\gamma^2-1)}\Bigg(4\gamma^4-12\gamma^2-3+C^{\rm ST}_1\Bigg)
        \nn\\
        &-\frac{2 m_1^2 m_2^2\gamma}{3(\gamma^2-1)^{5/2}}\Bigg(\Big(20\gamma^6-90\gamma^4+120\gamma^2-53\Big)+C_2^{\rm ST,1}\Bigg)
        \nn\\
        &-\frac{2m_1m_2^3}{(\gamma^2-1)^{5/2}}\Bigg(\Big(16\gamma^6-32\gamma^4+16\gamma^2-1\Big)+C_3^{\rm ST,1}\Bigg)\Bigg]+1\leftrightarrow2~,
\end{align}
where
\begin{align}
    C_1^{\rm ST} = &-\frac{1}{16} \alpha_1 \alpha_2 \Big(4 \tilde{d}_1^{(1)} \tilde{f}_1^{(1)} + 2 (\tilde{d}_2^{(1)}+          \tilde{f}_2^{(1)})-c_1 \alpha_1\Big) \Big( 4 \tilde{d}_1^{(2)} \tilde{f}_1^{(2)} +2 (\tilde{d}_2^{(2)}+\tilde{f}_2^{(2)})-c_1\alpha_2 \Big)
        \nn\\
    &+\frac{1}{4}\alpha_1 \alpha_2 \Big(8 + \alpha_1 \alpha_2+ 40 \gamma^2  \Big)~,
    \\
    C_2^{\rm ST,a} =&\frac{3}{2}\alpha_1\alpha_2\Big(50\gamma^4-102\gamma^2+49\Big)-\frac{3}{8}\alpha_1^2\alpha_2^2(2+\alpha_1\alpha_2+4\gamma^2)
    \nn\\
    &-\frac{1}{2}\alpha_{\cancel{a}}^2(\gamma^2-1)^2\Big(4\gamma^2+2+3\alpha_1\alpha_2\Big)
    \nn\\
    &-\frac{3}{8}\alpha_{\cancel{a}}^2(\gamma^2-1)(4\gamma^2-2+\alpha_1\alpha_2)\Big(4\tilde{d}_1^{({a})}\tilde{f}_1^{({a})}+2(\tilde{d}_2^{({a})}+\tilde{f}_2^{({a})})-c_1\alpha_{{a}}\Big)
         ,\\
    C_3^{\rm ST,a} = &\frac{1}{6}\alpha_1\alpha_2\big(20\gamma^4-46\gamma^2+17\big)-\frac{1}{8}\alpha_1^2\alpha_2^2(2+\alpha_1\alpha_2+4\gamma^2)
        \nn\\
        &-\frac{1}{3}\alpha_{\cancel{a}}^2(\gamma^2-1)^2(4\gamma^2+\alpha_1\alpha_2-2)
        \nn\\
        &+\frac{1}{8}\alpha_{\cancel{a}}^2(\gamma^2-1)\Big(2\big(\tilde{d}_2^{(a)}+\tilde{f}_2^{(a)}+2\tilde{d}_1^{(a)}\tilde{f}  _1^{(a)}\big)-c_1 \alpha_a\Big)\Big(6-8\gamma^2+\alpha_1\alpha_2(2\gamma^2-3)\Big)
        \nn\\
        &
        +\frac{1}{12}\alpha_{\cancel{a}}^3(\gamma^2-1)^2\Bigg(\alpha_a\big(2c_1^2-c_2\big)+2\big(\tilde{d}_3^{(a)}+\tilde{f}_3^{(a)}-3 \tilde{f}_1^{(a)} \tilde{f}_2^{(a)}\big) 
        \nn\\
        &-6 \tilde{d}_1^{(a)}\big(\tilde{d}_2^{(a)} + 2 \alpha_a\tilde{f}_1^{(a)} \big) -3c_1\big(\tilde{d}_2^{(a)} + \tilde{f}_2^{(a)} + 2 \tilde{d}_1^{(a)} \tilde{f}_1^{(a)} - \alpha_a^2\big) \Bigg)~,
\end{align}
with $\alpha_a =\tilde{d}_1^{(a)}+\tilde{f}_1^{(a)}$ and $\cancel{a}$ denoting the other particle ($\cancel{1} = 2$ and $\cancel{2} = 1$). The velocity components can be obtained straightforwardly by using the impulse at 1PM and 2PM into Eq.~\eqref{eq:iterR}. Finally, the scattering angle at 3PM order, obtained by inserting the above impulse into \eqref{eq:scat} is,

\begin{align}
        \frac{\chi_b^{(3)}}{\Gamma} = &\frac{1}{(\gamma^2-1)^{3/2}} \Bigg[-8\nu\ \sinh^{-1}\sqrt\frac{\gamma-1}{2}\Bigg(4\gamma^4-12\gamma^2-3+C^{\rm ST}_1\Bigg)
        \nn\\
        &+\frac{2\nu\gamma}{3(\gamma^2-1)^{3/2}}\Bigg(-2(\gamma^2-1)^2\Big(14\gamma^2+25\Big)+\frac{C_2^{\rm ST,1}+C_2^{\rm ST,2}}{2}\Bigg)
        \nn\\
        &+\frac{1}{(\gamma^2-1)^{3/2}}\Bigg(\left(\frac{m_2}{M}-\nu\right)C_3^{\rm ST,1}+\left(\frac{m_1}{M}-\nu\right)C_3^{\rm ST,2}\Bigg)
        \nn\\
        &+\frac{1+2\nu(\gamma-1)}{3(\gamma^2-1)^{3/2}}\Bigg(\Big(64\gamma^6-120\gamma^4+60\gamma^2-5\Big)+C_4^{\rm ST}\Bigg)\Bigg]~,
\end{align}
where
\begin{equation}
    C_4^{\rm ST}=\frac{(\alpha_1\alpha_2)^3}{4}+3\alpha_1\alpha_2(2\gamma^2-1)^2+\frac{3}{2}(\alpha_1\alpha_2)^2(2\gamma^2-1)~.
\end{equation}

\section{Discussions}
\label{sec:discuss}

In this work, we have extended the PM-EFT analysis to massless scalar-tensor theories up to 3PM order. Within the EFT framework, we computed the conservative dynamics and the corresponding impulse by explicitly solving the equations of motion to derive the corrections to the unperturbed solutions. We then discussed the decomposition of the impulse into its components proportional to the impact parameter and the velocities. From the impulse, we derived the scattering angle, which is the main result of our work.

To verify our PM results, we compared our expressions with the PN results of~\cite{Jain:2023vlf}. In particular, we expand the PM results in the small velocity regime and translate the scalar-tensor coefficients \eqref{eq:coeff11}-\eqref{eq:coeff12} into the conventions of Refs.~\cite{Damour:1995kt,Bernard:2018hta,Jain:2023vlf},
\begin{align}
    \alpha_a =&\ \tilde{d}_1^{(a)}+\tilde{f}_1^{(a)}~,\\ 
    \bar{\gamma} =&\ -2\frac{\alpha_1\alpha_2}{2+\alpha_1 \alpha_2}~,\\
    \delta_a =&\ 2\frac{\alpha_a^2}{(2+\alpha_1\alpha_2)^2}~,\\
\frac{\gamma_{AB}^2+4\beta_a}{\delta_{\cancel{a}}} =&\ 4 \tilde{d}_1^{(a)} \tilde{f}_1^{(a)} + 2 (\tilde{d}_2^{(a)}+          \tilde{f}_2^{(a)})-c_1 \big(\tilde{d}_1^{(a)}+\tilde{f}_1^{(a)}\big)~,\\
    -\frac{2\alpha_{a}}{\gamma_{AB}\delta_{\cancel{a}}}\epsilon_a=&\ 4\alpha_a^3+\alpha_a\big(2c_1^2-c_2\big)+2\big(\tilde{d}_3^{(a)}+\tilde{f}_3^{(a)}-3 \tilde{f}_1^{(a)} \tilde{f}_2^{(a)}\big) 
        \nn\\
        &-6 \tilde{d}_1^{(a)}\big(\tilde{d}_2^{(a)} + 2 \alpha_a\tilde{f}_1^{(a)} \big) -3c_1\big(\tilde{d}_2^{(a)} + \tilde{f}_2^{(a)} + 2 \tilde{d}_1^{(a)} \tilde{f}_1^{(a)} - \alpha_a^2\big).
\end{align}
Using the above translation, we find that our PM results are in complete agreement with the PN results of~\cite{Jain:2023vlf} obtained using the effective-one-body Hamiltonian up to $\mathcal{O}(1/j^3)$. We note that the modifications to the 3PN Lagrangian computed in~\cite{Bernard:inprep} do not affect the scattering angle at $\mathcal{O}(1/j^3)$, thus ensuring the consistency of the comparison\footnote{We have explicitly verified that changes in the 3PN Lagrangian only modifies the 3PN effective-one-body metric potential A, which, as shown in\cite{Jain:2023vlf} enters in the scattering angle at $\mathcal{O}(1/j^4)$.}. Our results also agree with the PM results of~\cite{Wilson-Gerow:2025xhr} when higher-derivative scalar tensor terms are neglected. As another consistency check of our results, we consider the binary black hole limit, under the assumption that the sensitivity of an isolated black hole holds for the binary black hole systems as well. For isolated black holes $s_a^{(0)} = 1/2$, upon substituting this value into our results of impulse (and scattering angle) we find that our results are indistinguishable from GR.

These results should be considered as a step forward in applying EFT techniques to modified theories of gravity for constructing waveform templates by translating the results into an effective-one-body description. As in GR, the radiation-reaction effects in the scattering angle start at the 3PM order in scalar-tensor theories~\cite{Damour:2020tta,Mougiakakos:2021ckm,Jakobsen:2021smu,Riva:2021vnj}. However, in scalar–tensor theories, in addition to graviton emission, a scalar mode may also be radiated, which could potentially lead to observable deviation in the phase of the waveform and hence might be essential for reliable waveform constructions.

\begin{acknowledgments}

T. J. is supported by the LabEx Junior Research Chair Fellowship at École normale supérieure, Paris. T. J. thanks Thibault Damour for useful discussions during the preparation of this manuscript.
LB and SM acknowledges financial support from the ANR PRoGRAM project, grant ANR-21-CE31-0003-001 and the EU Horizon 2020 Research
and Innovation Programme under the Marie Sklodowska-Curie Grant Agreement no.101007855.
This research was supported in part by Perimeter Institute for Theoretical Physics. Research at Perimeter Institute is supported in part by the Government of Canada through the Department of Innovation, Science and Economic Development and by the Province of Ontario through the Ministry of Colleges and Universities.

\end{acknowledgments}

\appendix

\section{Feynman diagrams at 3PM order}
\label{app:Feynman3PM}
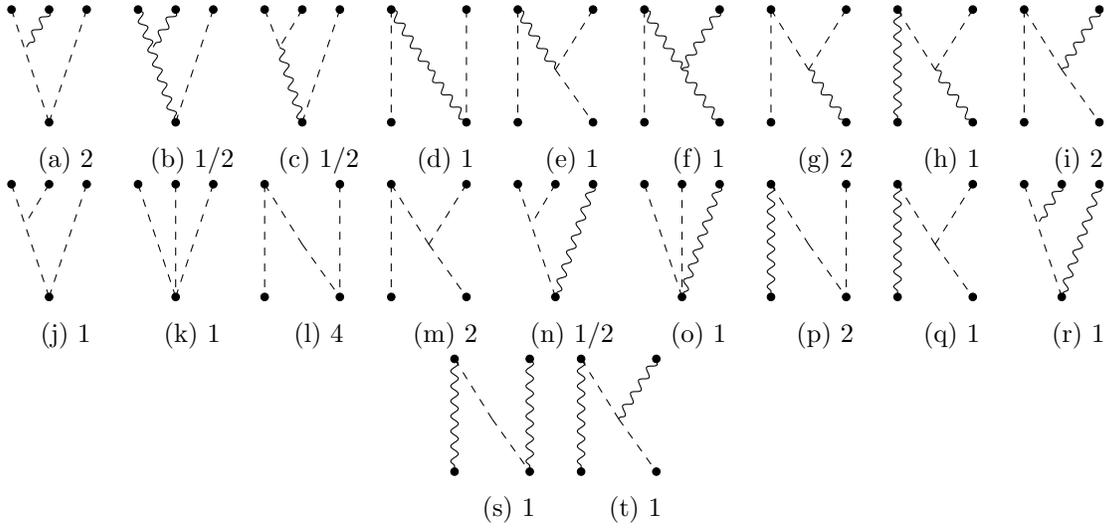
\begin{figure}[h]
    \centering
\begin{subfigure}[b]{0.1\textwidth}
    \begin{tikzpicture}
      [scale=1]
      \draw[scalar] (-0.5,0) -- (0,-1.5);
      \draw[boson] (0,0) -- (-0.3,-0.5);
      \draw[scalar] (0.5,0) -- (0,-1.5);
      \filldraw (-0.5,0) circle (0.05);
      \filldraw (0.5,0) circle (0.05);
      \filldraw (0,0) circle (0.05);
      \filldraw (0,-1.5) circle (0.05);
    \end{tikzpicture}
    \caption{2}
  \end{subfigure}
  \begin{subfigure}[b]{0.1\textwidth}
    \begin{tikzpicture}
      [scale=1]
      \draw[boson] (-0.5,0) -- (0,-1.5);
      \draw[boson] (0,0) -- (-0.3,-0.5);
      \draw[scalar] (0.5,0) -- (0,-1.5);
      \filldraw (-0.5,0) circle (0.05);
      \filldraw (0.5,0) circle (0.05);
      \filldraw (0,0) circle (0.05);
      \filldraw (0,-1.5) circle (0.05);
    \end{tikzpicture}
    \caption{1/2}
  \end{subfigure}
  \begin{subfigure}[b]{0.1\textwidth}
    \begin{tikzpicture}
      [scale=1]
      \draw[scalar] (-0.5,0) -- (-0.3,-0.5);
      \draw[boson] (-0.3,-0.5) -- (0,-1.5);
      \draw[scalar] (0,0) -- (-0.3,-0.5);
      \draw[scalar] (0.5,0) -- (0,-1.5);
      \filldraw (-0.5,0) circle (0.05);
      \filldraw (0.5,0) circle (0.05);
      \filldraw (0,0) circle (0.05);
      \filldraw (0,-1.5) circle (0.05);
    \end{tikzpicture}
    \caption{1/2}
  \end{subfigure}
   \begin{subfigure}[b]{0.1\textwidth}
    \begin{tikzpicture}
      [scale=1]
      \draw[boson] (-0.5,0) -- (0,-0.8);
      \draw[scalar] (0.5,0) -- (0.5,-1.5);
      \draw[scalar] (-0.5,0) -- (-0.5,-1.5);
      \draw[boson] (0,-0.8) -- (0.5,-1.5);
      \filldraw (-0.5,0) circle (0.05);
      \filldraw (0.5,0) circle (0.05);
      \filldraw (-0.5,-1.5) circle (0.05);
      \filldraw (0.5,-1.5) circle (0.05);
    \end{tikzpicture}
    \caption{1}
  \end{subfigure}
  \begin{subfigure}[b]{0.1\textwidth}
    \begin{tikzpicture}
      [scale=1]
      \draw[boson] (-0.5,0) -- (0,-0.8);
      \draw[scalar] (0.5,0) -- (0,-0.8);
      \draw[scalar] (-0.5,0) -- (-0.5,-1.5);
      \draw[scalar] (0,-0.8) -- (0.5,-1.5);
      \filldraw (-0.5,0) circle (0.05);
      \filldraw (0.5,0) circle (0.05);
      \filldraw (-0.5,-1.5) circle (0.05);
      \filldraw (0.5,-1.5) circle (0.05);
    \end{tikzpicture}
    \caption{1}
  \end{subfigure}
  \begin{subfigure}[b]{0.1\textwidth}
    \begin{tikzpicture}
      [scale=1]
      \draw[boson] (-0.5,0) -- (0,-0.8);
      \draw[boson] (0.5,0) -- (0,-0.8);
      \draw[scalar] (-0.5,0) -- (-0.5,-1.5);
      \draw[boson] (0,-0.8) -- (0.5,-1.5);
      \filldraw (-0.5,0) circle (0.05);
      \filldraw (0.5,0) circle (0.05);
      \filldraw (-0.5,-1.5) circle (0.05);
      \filldraw (0.5,-1.5) circle (0.05);
    \end{tikzpicture}
    \caption{1}
  \end{subfigure}
  \begin{subfigure}[b]{0.1\textwidth}
    \begin{tikzpicture}
      [scale=1]
      \draw[scalar] (-0.5,0) -- (0,-0.8);
      \draw[scalar] (0.5,0) -- (0,-0.8);
      \draw[scalar] (-0.5,0) -- (-0.5,-1.5);
      \draw[boson] (0,-0.8) -- (0.5,-1.5);
      \filldraw (-0.5,0) circle (0.05);
      \filldraw (0.5,0) circle (0.05);
      \filldraw (-0.5,-1.5) circle (0.05);
      \filldraw (0.5,-1.5) circle (0.05);
    \end{tikzpicture}
    \caption{2}
  \end{subfigure}
  \begin{subfigure}[b]{0.1\textwidth}
    \begin{tikzpicture}
      [scale=1]
      \draw[scalar] (-0.5,0) -- (0,-0.8);
      \draw[scalar] (0.5,0) -- (0,-0.8);
      \draw[boson] (-0.5,0) -- (-0.5,-1.5);
      \draw[boson] (0,-0.8) -- (0.5,-1.5);
      \filldraw (-0.5,0) circle (0.05);
      \filldraw (0.5,0) circle (0.05);
      \filldraw (-0.5,-1.5) circle (0.05);
      \filldraw (0.5,-1.5) circle (0.05);
    \end{tikzpicture}
    \caption{1}
  \end{subfigure}
  \begin{subfigure}[b]{0.1\textwidth}
    \begin{tikzpicture}
      [scale=1]
      \draw[scalar] (-0.5,0) -- (0,-0.8);
      \draw[boson] (0.5,0) -- (0,-0.8);
      \draw[scalar] (-0.5,0) -- (-0.5,-1.5);
      \draw[scalar] (0,-0.8) -- (0.5,-1.5);
      \filldraw (-0.5,0) circle (0.05);
      \filldraw (0.5,0) circle (0.05);
      \filldraw (-0.5,-1.5) circle (0.05);
      \filldraw (0.5,-1.5) circle (0.05);
    \end{tikzpicture}
    \caption{2}
  \end{subfigure}
\begin{subfigure}[b]{0.1\textwidth}
    \begin{tikzpicture}
      [scale=1]
      \draw[scalar] (-0.5,0) -- (0,-1.5);
      \draw[scalar] (0,0) -- (-0.3,-0.5);
      \draw[scalar] (0.5,0) -- (0,-1.5);
      \filldraw (-0.5,0) circle (0.05);
      \filldraw (0.5,0) circle (0.05);
      \filldraw (0,0) circle (0.05);
      \filldraw (0,-1.5) circle (0.05);
    \end{tikzpicture}
    \caption{1}
  \end{subfigure}
  \begin{subfigure}[b]{0.1\textwidth}
    \begin{tikzpicture}
      [scale=1]
      \draw[scalar] (-0.5,0) -- (0,-1.5);
      \draw[scalar] (0,0) -- (0,-1.5);
      \draw[scalar] (0.5,0) -- (0,-1.5);
      \filldraw (-0.5,0) circle (0.05);
      \filldraw (0.5,0) circle (0.05);
      \filldraw (0,0) circle (0.05);
      \filldraw (0,-1.5) circle (0.05);
    \end{tikzpicture}
    \caption{1}
  \end{subfigure}
 \begin{subfigure}[b]{0.1\textwidth}
    \begin{tikzpicture}
      [scale=1]
      \draw[scalar] (-0.5,0) -- (0,-0.8);
      \draw[scalar] (0.5,0) -- (0.5,-1.5);
      \draw[scalar] (-0.5,0) -- (-0.5,-1.5);
      \draw[scalar] (0,-0.8) -- (0.5,-1.5);
      \filldraw (-0.5,0) circle (0.05);
      \filldraw (0.5,0) circle (0.05);
      \filldraw (-0.5,-1.5) circle (0.05);
      \filldraw (0.5,-1.5) circle (0.05);
    \end{tikzpicture}
    \caption{4}
  \end{subfigure}
  \begin{subfigure}[b]{0.1\textwidth}
    \begin{tikzpicture}
      [scale=1]
      \draw[scalar] (-0.5,0) -- (0,-0.8);
      \draw[scalar] (0.5,0) -- (0,-0.8);
      \draw[scalar] (-0.5,0) -- (-0.5,-1.5);
      \draw[scalar] (0,-0.8) -- (0.5,-1.5);
      \filldraw (-0.5,0) circle (0.05);
      \filldraw (0.5,0) circle (0.05);
      \filldraw (-0.5,-1.5) circle (0.05);
      \filldraw (0.5,-1.5) circle (0.05);
    \end{tikzpicture}
    \caption{2}
  \end{subfigure}
  \begin{subfigure}[b]{0.1\textwidth}
    \begin{tikzpicture}
      [scale=1]
      \draw[scalar] (-0.5,0) -- (0,-1.5);
      \draw[scalar] (0,0) -- (-0.3,-0.5);
      \draw[boson] (0.5,0) -- (0,-1.5);
      \filldraw (-0.5,0) circle (0.05);
      \filldraw (0.5,0) circle (0.05);
      \filldraw (0,0) circle (0.05);
      \filldraw (0,-1.5) circle (0.05);
    \end{tikzpicture}
    \caption{1/2}
  \end{subfigure}
  \begin{subfigure}[b]{0.1\textwidth}
    \begin{tikzpicture}
      [scale=1]
      \draw[scalar] (-0.5,0) -- (0,-1.5);
      \draw[scalar] (0,0) -- (0,-1.5);
      \draw[boson] (0.5,0) -- (0,-1.5);
      \filldraw (-0.5,0) circle (0.05);
      \filldraw (0.5,0) circle (0.05);
      \filldraw (0,0) circle (0.05);
      \filldraw (0,-1.5) circle (0.05);
    \end{tikzpicture}
    \caption{1}
  \end{subfigure}
 \begin{subfigure}[b]{0.1\textwidth}
    \begin{tikzpicture}
      [scale=1]
      \draw[scalar] (-0.5,0) -- (0,-0.8);
      \draw[scalar] (0.5,0) -- (0.5,-1.5);
      \draw[boson] (-0.5,0) -- (-0.5,-1.5);
      \draw[scalar] (0,-0.8) -- (0.5,-1.5);
      \filldraw (-0.5,0) circle (0.05);
      \filldraw (0.5,0) circle (0.05);
      \filldraw (-0.5,-1.5) circle (0.05);
      \filldraw (0.5,-1.5) circle (0.05);
    \end{tikzpicture}
    \caption{2}
  \end{subfigure}
  \begin{subfigure}[b]{0.1\textwidth}
    \begin{tikzpicture}
      [scale=1]
      \draw[scalar] (-0.5,0) -- (0,-0.8);
      \draw[scalar] (0.5,0) -- (0,-0.8);
      \draw[boson] (-0.5,0) -- (-0.5,-1.5);
      \draw[scalar] (0,-0.8) -- (0.5,-1.5);
      \filldraw (-0.5,0) circle (0.05);
      \filldraw (0.5,0) circle (0.05);
      \filldraw (-0.5,-1.5) circle (0.05);
      \filldraw (0.5,-1.5) circle (0.05);
    \end{tikzpicture}
    \caption{1}
  \end{subfigure}
    \begin{subfigure}[b]{0.1\textwidth}
    \begin{tikzpicture}
      [scale=1]
      \draw[scalar] (-0.5,0) -- (0,-1.5);
      \draw[boson] (0,0) -- (-0.3,-0.5);
      \draw[boson] (0.5,0) -- (0,-1.5);
      \filldraw (-0.5,0) circle (0.05);
      \filldraw (0.5,0) circle (0.05);
      \filldraw (0,0) circle (0.05);
      \filldraw (0,-1.5) circle (0.05);
    \end{tikzpicture}
    \caption{1}
  \end{subfigure}
  \begin{subfigure}[b]{0.1\textwidth}
    \begin{tikzpicture}
      [scale=1]
      \draw[scalar] (-0.5,0) -- (0,-0.8);
      \draw[boson] (0.5,0) -- (0.5,-1.5);
      \draw[boson] (-0.5,0) -- (-0.5,-1.5);
      \draw[scalar] (0,-0.8) -- (0.5,-1.5);
      \filldraw (-0.5,0) circle (0.05);
      \filldraw (0.5,0) circle (0.05);
      \filldraw (-0.5,-1.5) circle (0.05);
      \filldraw (0.5,-1.5) circle (0.05);
    \end{tikzpicture}
    \caption{1}
  \end{subfigure}
  \begin{subfigure}[b]{0.1\textwidth}
    \begin{tikzpicture}
      [scale=1]
      \draw[scalar] (-0.5,0) -- (0,-0.8);
      \draw[boson] (0.5,0) -- (0,-0.8);
      \draw[boson] (-0.5,0) -- (-0.5,-1.5);
      \draw[scalar] (0,-0.8) -- (0.5,-1.5);
      \filldraw (-0.5,0) circle (0.05);
      \filldraw (0.5,0) circle (0.05);
      \filldraw (-0.5,-1.5) circle (0.05);
      \filldraw (0.5,-1.5) circle (0.05);
    \end{tikzpicture}
    \caption{1}
  \end{subfigure}
      \caption{Feynman diagrams due to worldline nonlinearities at 3PM order.}
    \label{fig:worldline}
  \end{figure}

In this Appendix, we list all the Feynman diagrams required at 3PM order due to the four graviton bulk vertices in Fig.~\ref{fig:bulk4graviton}, the four scalar bulk vertices in Fig.~\ref{fig:bulk4scalar}, the three-scalar one-graviton bulk vertices in Fig.~\ref{fig:bulk3scalar}, and the two-scalar two-graviton bulk vertices in Fig.~\ref{fig:bulk2scalar}, along with their symmetry factors. The diagrams due to nonlinearities of the worldline action are given in Fig.~\ref{fig:worldline}.
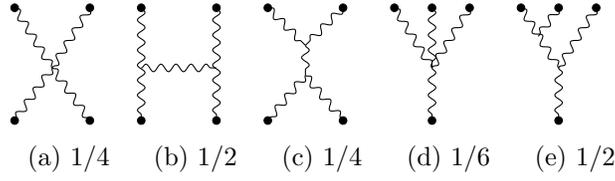
\begin{figure}[h]
  \centering
    \begin{subfigure}[b]{0.1\textwidth}
    \begin{tikzpicture}
      [scale=1]
      \draw[boson] (-0.5,0) -- (0,-0.8);
      \draw[boson] (0.5,0) -- (0,-0.8);
      \draw[boson] (0,-0.8) -- (-0.5,-1.5);
      \draw[boson] (0,-0.8) -- (0.5,-1.5);
      \filldraw (-0.5,0) circle (0.05);
      \filldraw (0.5,0) circle (0.05);
      \filldraw (-0.5,-1.5) circle (0.05);
      \filldraw (0.5,-1.5) circle (0.05);
    \end{tikzpicture}
    \caption{1/4}
  \end{subfigure}
   \begin{subfigure}[b]{0.1\textwidth}
    \begin{tikzpicture}
      [scale=1]
      \draw[boson] (-0.5,0) -- (-0.5,-1.5);
      \draw[boson] (0.5,0) -- (0.5,-1.5);
      \draw[boson] (-0.5,-0.8) -- (0.5,-0.8);
      \filldraw (-0.5,0) circle (0.05);
      \filldraw (0.5,0) circle (0.05);
      \filldraw (-0.5,-1.5) circle (0.05);
      \filldraw (0.5,-1.5) circle (0.05);
    \end{tikzpicture}
    \caption{1/2}
  \end{subfigure}
    \begin{subfigure}[b]{0.1\textwidth}
    \begin{tikzpicture}
      [scale=1]
      \draw[boson] (-0.5,0) -- (0,-0.5);
      \draw[boson] (0.5,0) -- (0,-0.5);
      \draw[boson] (0,-0.5) -- (0,-0.9);
      \draw[boson] (0,-0.9) -- (-0.5,-1.5);
      \draw[boson] (0,-0.9) -- (0.5,-1.5);
      \filldraw (-0.5,0) circle (0.05);
      \filldraw (0.5,0) circle (0.05);
      \filldraw (-0.5,-1.5) circle (0.05);
      \filldraw (0.5,-1.5) circle (0.05);
    \end{tikzpicture}
    \caption{1/4}
  \end{subfigure}
    \begin{subfigure}[b]{0.1\textwidth}
    \begin{tikzpicture}
      [scale=1]
      \draw[boson] (-0.5,0) -- (0,-0.8);
      \draw[boson] (0,0) -- (0,-0.8);
      \draw[boson] (0.5,0) -- (0,-0.8);
      \draw[boson] (0,-0.8) -- (0,-1.5);
      \filldraw (-0.5,0) circle (0.05);
      \filldraw (0.5,0) circle (0.05);
      \filldraw (0,0) circle (0.05);
      \filldraw (0,-1.5) circle (0.05);
    \end{tikzpicture}
    \caption{1/6}
  \end{subfigure}
    \begin{subfigure}[b]{0.1\textwidth}
    \begin{tikzpicture}
      [scale=1]
      \draw[boson] (-0.5,0) -- (0,-0.8);
      \draw[boson] (0,0) -- (-0.27,-0.38);
      \draw[boson] (0.5,0) -- (0,-0.8);
      \draw[boson] (0,-0.8) -- (0,-1.5);
      \filldraw (-0.5,0) circle (0.05);
      \filldraw (0.5,0) circle (0.05);
      \filldraw (0,0) circle (0.05);
      \filldraw (0,-1.5) circle (0.05);
    \end{tikzpicture}
    \caption{1/2}
  \end{subfigure}
    \caption{Feynman diagrams due to four-scalar bulk vertices.}
    \label{fig:bulk4graviton}
  \end{figure}
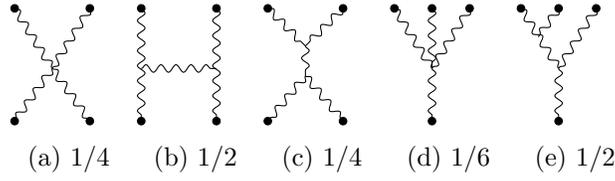
\begin{figure}[h]
  \centering
    \begin{subfigure}[b]{0.1\textwidth}
    \begin{tikzpicture}
      [scale=1]
      \draw[boson] (-0.5,0) -- (0,-0.8);
      \draw[boson] (0.5,0) -- (0,-0.8);
      \draw[boson] (0,-0.8) -- (-0.5,-1.5);
      \draw[boson] (0,-0.8) -- (0.5,-1.5);
      \filldraw (-0.5,0) circle (0.05);
      \filldraw (0.5,0) circle (0.05);
      \filldraw (-0.5,-1.5) circle (0.05);
      \filldraw (0.5,-1.5) circle (0.05);
    \end{tikzpicture}
    \caption{1/4}
  \end{subfigure}
   \begin{subfigure}[b]{0.1\textwidth}
    \begin{tikzpicture}
      [scale=1]
      \draw[boson] (-0.5,0) -- (-0.5,-1.5);
      \draw[boson] (0.5,0) -- (0.5,-1.5);
      \draw[boson] (-0.5,-0.8) -- (0.5,-0.8);
      \filldraw (-0.5,0) circle (0.05);
      \filldraw (0.5,0) circle (0.05);
      \filldraw (-0.5,-1.5) circle (0.05);
      \filldraw (0.5,-1.5) circle (0.05);
    \end{tikzpicture}
    \caption{1/2}
  \end{subfigure}
    \begin{subfigure}[b]{0.1\textwidth}
    \begin{tikzpicture}
      [scale=1]
      \draw[boson] (-0.5,0) -- (0,-0.5);
      \draw[boson] (0.5,0) -- (0,-0.5);
      \draw[boson] (0,-0.5) -- (0,-0.9);
      \draw[boson] (0,-0.9) -- (-0.5,-1.5);
      \draw[boson] (0,-0.9) -- (0.5,-1.5);
      \filldraw (-0.5,0) circle (0.05);
      \filldraw (0.5,0) circle (0.05);
      \filldraw (-0.5,-1.5) circle (0.05);
      \filldraw (0.5,-1.5) circle (0.05);
    \end{tikzpicture}
    \caption{1/4}
  \end{subfigure}
    \begin{subfigure}[b]{0.1\textwidth}
    \begin{tikzpicture}
      [scale=1]
      \draw[boson] (-0.5,0) -- (0,-0.8);
      \draw[boson] (0,0) -- (0,-0.8);
      \draw[boson] (0.5,0) -- (0,-0.8);
      \draw[boson] (0,-0.8) -- (0,-1.5);
      \filldraw (-0.5,0) circle (0.05);
      \filldraw (0.5,0) circle (0.05);
      \filldraw (0,0) circle (0.05);
      \filldraw (0,-1.5) circle (0.05);
    \end{tikzpicture}
    \caption{1/6}
  \end{subfigure}
    \begin{subfigure}[b]{0.1\textwidth}
    \begin{tikzpicture}
      [scale=1]
      \draw[boson] (-0.5,0) -- (0,-0.8);
      \draw[boson] (0,0) -- (-0.27,-0.38);
      \draw[boson] (0.5,0) -- (0,-0.8);
      \draw[boson] (0,-0.8) -- (0,-1.5);
      \filldraw (-0.5,0) circle (0.05);
      \filldraw (0.5,0) circle (0.05);
      \filldraw (0,0) circle (0.05);
      \filldraw (0,-1.5) circle (0.05);
    \end{tikzpicture}
    \caption{1/2}
  \end{subfigure}
    \caption{Feynman diagrams due to four-graviton bulk vertices.}
    \label{fig:bulk4scalar}
  \end{figure}
  \begin{figure}[h]
  \centering
   \begin{subfigure}[b]{0.1\textwidth}
    \begin{tikzpicture}
      [scale=1]
      \draw[scalar] (-0.5,0) -- (0,-0.8);
      \draw[scalar] (0.5,0) -- (0,-0.8);
      \draw[scalar] (0,-0.8) -- (-0.5,-1.5);
      \draw[boson] (0,-0.8) -- (0.5,-1.5);
      \filldraw (-0.5,0) circle (0.05);
      \filldraw (0.5,0) circle (0.05);
      \filldraw (-0.5,-1.5) circle (0.05);
      \filldraw (0.5,-1.5) circle (0.05);
    \end{tikzpicture}
    \caption{1/2}
  \end{subfigure}
    \begin{subfigure}[b]{0.1\textwidth}
    \begin{tikzpicture}
      [scale=1]
      \draw[scalar] (-0.5,0) -- (-0.5,-0.8);
      \draw[boson]  (-0.5,-0.8) -- (-0.5,-1.5);
      \draw[scalar] (0.5,0) -- (0.5,-1.5);
      \draw[scalar] (-0.5,-0.8) -- (0.5,-0.8);
      \filldraw (-0.5,0) circle (0.05);
      \filldraw (0.5,0) circle (0.05);
      \filldraw (-0.5,-1.5) circle (0.05);
      \filldraw (0.5,-1.5) circle (0.05);
    \end{tikzpicture}
    \caption{1}
  \end{subfigure}
     \begin{subfigure}[b]{0.1\textwidth}
    \begin{tikzpicture}
      [scale=1]
      \draw[scalar] (-0.5,0) -- (0,-0.5);
      \draw[scalar] (0.5,0) -- (0,-0.5);
      \draw[scalar] (0,-0.5) -- (0,-0.9);
      \draw[scalar] (0,-0.9) -- (-0.5,-1.5);
      \draw[boson] (0,-0.9) -- (0.5,-1.5);
      \filldraw (-0.5,0) circle (0.05);
      \filldraw (0.5,0) circle (0.05);
      \filldraw (-0.5,-1.5) circle (0.05);
      \filldraw (0.5,-1.5) circle (0.05);
    \end{tikzpicture}
    \caption{1/2}
  \end{subfigure}
    \begin{subfigure}[b]{0.1\textwidth}
    \begin{tikzpicture}
      [scale=1]
      \draw[scalar] (-0.5,0) -- (0,-0.8);
      \draw[scalar] (0,0) -- (0,-0.8);
      \draw[scalar] (0.5,0) -- (0,-0.8);
      \draw[boson] (0,-0.8) -- (0,-1.5);
      \filldraw (-0.5,0) circle (0.05);
      \filldraw (0.5,0) circle (0.05);
      \filldraw (0,0) circle (0.05);
      \filldraw (0,-1.5) circle (0.05);
    \end{tikzpicture}
    \caption{1/6}
  \end{subfigure}
    \begin{subfigure}[b]{0.1\textwidth}
    \begin{tikzpicture}
      [scale=1]
      \draw[scalar] (-0.5,0) -- (0,-0.8);
      \draw[scalar] (0,0) -- (0,-0.8);
      \draw[boson] (0.5,0) -- (0,-0.8);
      \draw[scalar] (0,-0.8) -- (0,-1.5);
      \filldraw (-0.5,0) circle (0.05);
      \filldraw (0.5,0) circle (0.05);
      \filldraw (0,0) circle (0.05);
      \filldraw (0,-1.5) circle (0.05);
    \end{tikzpicture}
    \caption{1/2}
  \end{subfigure}
  \begin{subfigure}[b]{0.1\textwidth}
    \begin{tikzpicture}
      [scale=1]
      \draw[scalar] (-0.5,0) -- (0,-0.8);
      \draw[scalar] (0,0) -- (-0.29,-0.38);
      \draw[scalar] (0.5,0) -- (0,-0.8);
      \draw[boson] (0,-0.8) -- (0,-1.5);
      \filldraw (-0.5,0) circle (0.05);
      \filldraw (0.5,0) circle (0.05);
      \filldraw (0,0) circle (0.05);
      \filldraw (0,-1.5) circle (0.05);
    \end{tikzpicture}
    \caption{1/2}
  \end{subfigure}
   \begin{subfigure}[b]{0.1\textwidth}
    \begin{tikzpicture}
      [scale=1]
      \draw[scalar] (-0.5,0) -- (0,-0.8);
      \draw[scalar] (0,0) -- (-0.29,-0.38);
      \draw[boson] (0.5,0) -- (0,-0.8);
      \draw[scalar] (0,-0.8) -- (0,-1.5);
      \filldraw (-0.5,0) circle (0.05);
      \filldraw (0.5,0) circle (0.05);
      \filldraw (0,0) circle (0.05);
      \filldraw (0,-1.5) circle (0.05);
    \end{tikzpicture}
    \caption{1/2}
  \end{subfigure}
       \begin{subfigure}[b]{0.1\textwidth}
    \centering
    \begin{tikzpicture}
      [scale=1]
      \draw[scalar] (-0.5,0) -- (-0.5,-1.5);
      \draw[scalar] (0.5,0) -- (0.5,-1.5);
      \draw[boson] (-0.5,-0.8) -- (0.5,-0.8);
      \filldraw (-0.5,0) circle (0.05);
      \filldraw (0.5,0) circle (0.05);
      \filldraw (-0.5,-1.5) circle (0.05);
      \filldraw (0.5,-1.5) circle (0.05);
    \end{tikzpicture}
    \caption{1/2}
  \end{subfigure}
    \begin{subfigure}[b]{0.1\textwidth}
    \begin{tikzpicture}
      [scale=1]
      \draw[scalar] (-0.5,0) -- (0,-0.5);
      \draw[scalar] (0.5,0) -- (0,-0.5);
      \draw[boson] (0,-0.5) -- (0,-0.9);
      \draw[scalar] (0,-0.9) -- (-0.5,-1.5);
      \draw[scalar] (0,-0.9) -- (0.5,-1.5);
      \filldraw (-0.5,0) circle (0.05);
      \filldraw (0.5,0) circle (0.05);
      \filldraw (-0.5,-1.5) circle (0.05);
      \filldraw (0.5,-1.5) circle (0.05);
    \end{tikzpicture}
    \caption{1/4}
  \end{subfigure}
    \begin{subfigure}[b]{0.1\textwidth}
    \begin{tikzpicture}
      [scale=1]
      \draw[scalar] (-0.5,0) -- (-0.2,-0.5);
      \draw[boson] (-0.2,-0.5) -- (0,-0.8);
      \draw[scalar] (0,0) -- (-0.29,-0.38);
      \draw[scalar] (0.5,0) -- (0,-0.8);
      \draw[scalar] (0,-0.8) -- (0,-1.5);
      \filldraw (-0.5,0) circle (0.05);
      \filldraw (0.5,0) circle (0.05);
      \filldraw (0,0) circle (0.05);
      \filldraw (0,-1.5) circle (0.05);
    \end{tikzpicture}
    \caption{1/2}
  \end{subfigure}
  \begin{subfigure}[b]{0.1\textwidth}
    \begin{tikzpicture}
      [scale=1]
      \draw[scalar] (-0.5,0) -- (0,-0.8);
      \draw[boson] (0,0) -- (-0.29,-0.38);
      \draw[scalar] (0.5,0) -- (0,-0.8);
      \draw[scalar] (0,-0.8) -- (0,-1.5);
      \filldraw (-0.5,0) circle (0.05);
      \filldraw (0.5,0) circle (0.05);
      \filldraw (0,0) circle (0.05);
      \filldraw (0,-1.5) circle (0.05);
    \end{tikzpicture}
    \caption{1}
  \end{subfigure}
    \caption{Feynman diagrams due to three-scalar one-graviton bulk vertices.}
    \label{fig:bulk3scalar}
  \end{figure}
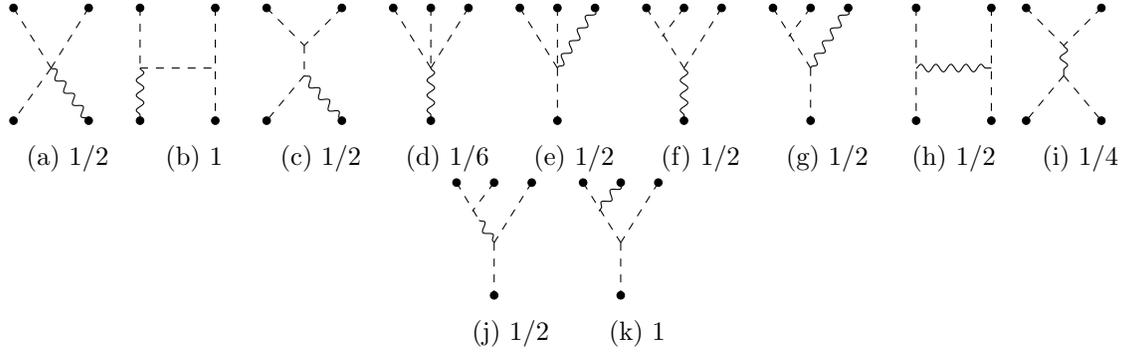
  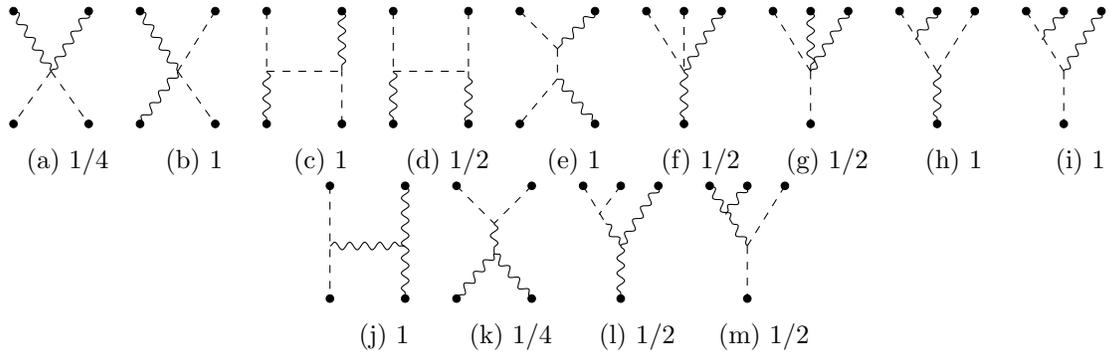
\begin{figure}[h]
  \centering
      \begin{subfigure}[b]{0.1\textwidth}
    \begin{tikzpicture}
      [scale=1]
      \draw[boson] (-0.5,0) -- (0,-0.8);
      \draw[boson] (0.5,0) -- (0,-0.8);
      \draw[scalar] (0,-0.8) -- (-0.5,-1.5);
      \draw[scalar] (0,-0.8) -- (0.5,-1.5);
      \filldraw (-0.5,0) circle (0.05);
      \filldraw (0.5,0) circle (0.05);
      \filldraw (-0.5,-1.5) circle (0.05);
      \filldraw (0.5,-1.5) circle (0.05);
    \end{tikzpicture}
    \caption{1/4}
  \end{subfigure} 
   \begin{subfigure}[b]{0.1\textwidth}
    \begin{tikzpicture}
      [scale=1]
      \draw[boson] (-0.5,0) -- (0,-0.8);
      \draw[scalar] (0.5,0) -- (0,-0.8);
      \draw[boson] (0,-0.8) -- (-0.5,-1.5);
      \draw[scalar] (0,-0.8) -- (0.5,-1.5);
      \filldraw (-0.5,0) circle (0.05);
      \filldraw (0.5,0) circle (0.05);
      \filldraw (-0.5,-1.5) circle (0.05);
      \filldraw (0.5,-1.5) circle (0.05);
    \end{tikzpicture}
    \caption{1}
  \end{subfigure}
  \begin{subfigure}[b]{0.1\textwidth}
    \begin{tikzpicture}
      [scale=1]
      \draw[scalar] (-0.5,0) -- (-0.5,-0.8);
      \draw[boson]  (-0.5,-0.8) -- (-0.5,-1.5);
      \draw[boson] (0.5,0) -- (0.5,-0.8);
      \draw[scalar] (0.5,-1.5) -- (0.5,-0.8);
      \draw[scalar] (-0.5,-0.8) -- (0.5,-0.8);
      \filldraw (-0.5,0) circle (0.05);
      \filldraw (0.5,0) circle (0.05);
      \filldraw (-0.5,-1.5) circle (0.05);
      \filldraw (0.5,-1.5) circle (0.05);
    \end{tikzpicture}
    \caption{1}
  \end{subfigure}
     \begin{subfigure}[b]{0.1\textwidth}
    \begin{tikzpicture}
      [scale=1]
      \draw[scalar] (-0.5,0) -- (-0.5,-0.8);
      \draw[boson]  (-0.5,-0.8) -- (-0.5,-1.5);
      \draw[scalar] (0.5,0) -- (0.5,-0.8);
      \draw[boson] (0.5,-1.5) -- (0.5,-0.8);
      \draw[scalar] (-0.5,-0.8) -- (0.5,-0.8);
      \filldraw (-0.5,0) circle (0.05);
      \filldraw (0.5,0) circle (0.05);
      \filldraw (-0.5,-1.5) circle (0.05);
      \filldraw (0.5,-1.5) circle (0.05);
    \end{tikzpicture}
    \caption{1/2}
  \end{subfigure}
  \begin{subfigure}[b]{0.1\textwidth}
    \begin{tikzpicture}
      [scale=1]
      \draw[scalar] (-0.5,0) -- (0,-0.5);
      \draw[boson] (0.5,0) -- (0,-0.5);
      \draw[scalar] (0,-0.5) -- (0,-0.9);
      \draw[scalar] (0,-0.9) -- (-0.5,-1.5);
      \draw[boson] (0,-0.9) -- (0.5,-1.5);
      \filldraw (-0.5,0) circle (0.05);
      \filldraw (0.5,0) circle (0.05);
      \filldraw (-0.5,-1.5) circle (0.05);
      \filldraw (0.5,-1.5) circle (0.05);
    \end{tikzpicture}
    \caption{1}
  \end{subfigure}
 \begin{subfigure}[b]{0.1\textwidth}
    \begin{tikzpicture}
      [scale=1]
      \draw[scalar] (-0.5,0) -- (0,-0.8);
      \draw[scalar] (0,0) -- (0,-0.8);
      \draw[boson] (0.5,0) -- (0,-0.8);
      \draw[boson] (0,-0.8) -- (0,-1.5);
      \filldraw (-0.5,0) circle (0.05);
      \filldraw (0.5,0) circle (0.05);
      \filldraw (0,0) circle (0.05);
      \filldraw (0,-1.5) circle (0.05);
    \end{tikzpicture}
    \caption{1/2}
  \end{subfigure}
  \begin{subfigure}[b]{0.1\textwidth}
    \begin{tikzpicture}
      [scale=1]
      \draw[scalar] (-0.5,0) -- (0,-0.8);
      \draw[boson] (0,0) -- (0,-0.8);
      \draw[boson] (0.5,0) -- (0,-0.8);
      \draw[scalar] (0,-0.8) -- (0,-1.5);
      \filldraw (-0.5,0) circle (0.05);
      \filldraw (0.5,0) circle (0.05);
      \filldraw (0,0) circle (0.05);
      \filldraw (0,-1.5) circle (0.05);
    \end{tikzpicture}
    \caption{1/2}
  \end{subfigure}
  \begin{subfigure}[b]{0.1\textwidth}
    \begin{tikzpicture}
      [scale=1]
      \draw[scalar] (-0.5,0) -- (0,-0.8);
      \draw[boson] (0,0) -- (-0.29,-0.38);
      \draw[scalar] (0.5,0) -- (0,-0.8);
      \draw[boson] (0,-0.8) -- (0,-1.5);
      \filldraw (-0.5,0) circle (0.05);
      \filldraw (0.5,0) circle (0.05);
      \filldraw (0,0) circle (0.05);
      \filldraw (0,-1.5) circle (0.05);
    \end{tikzpicture}
    \caption{1}
  \end{subfigure}
    \begin{subfigure}[b]{0.1\textwidth}
    \begin{tikzpicture}
      [scale=1]
      \draw[scalar] (-0.5,0) -- (0,-0.8);
      \draw[boson] (0,0) -- (-0.29,-0.38);
      \draw[boson] (0.5,0) -- (0,-0.8);
      \draw[scalar] (0,-0.8) -- (0,-1.5);
      \filldraw (-0.5,0) circle (0.05);
      \filldraw (0.5,0) circle (0.05);
      \filldraw (0,0) circle (0.05);
      \filldraw (0,-1.5) circle (0.05);
    \end{tikzpicture}
    \caption{1}
  \end{subfigure}
   \begin{subfigure}[b]{0.1\textwidth}
    \begin{tikzpicture}
      [scale=1]
      \draw[scalar] (-0.5,0) -- (-0.5,-1.5);
      \draw[boson] (0.5,0) -- (0.5,-1.5);
      \draw[boson] (-0.5,-0.8) -- (0.5,-0.8);
      \filldraw (-0.5,0) circle (0.05);
      \filldraw (0.5,0) circle (0.05);
      \filldraw (-0.5,-1.5) circle (0.05);
      \filldraw (0.5,-1.5) circle (0.05);
    \end{tikzpicture}
    \caption{1}
  \end{subfigure}
   \begin{subfigure}[b]{0.1\textwidth}
    \begin{tikzpicture}
      [scale=1]
      \draw[scalar] (-0.5,0) -- (0,-0.5);
      \draw[scalar] (0.5,0) -- (0,-0.5);
      \draw[boson] (0,-0.5) -- (0,-0.9);
      \draw[boson] (0,-0.9) -- (-0.5,-1.5);
      \draw[boson] (0,-0.9) -- (0.5,-1.5);
      \filldraw (-0.5,0) circle (0.05);
      \filldraw (0.5,0) circle (0.05);
      \filldraw (-0.5,-1.5) circle (0.05);
      \filldraw (0.5,-1.5) circle (0.05);
    \end{tikzpicture}
    \caption{1/4}
  \end{subfigure}
   \begin{subfigure}[b]{0.1\textwidth}
    \begin{tikzpicture}
      [scale=1]
      \draw[scalar] (-0.5,0) -- (-0.2,-0.5);
      \draw[boson] (-0.2,-0.5) -- (0,-0.8);
      \draw[scalar] (0,0) -- (-0.29,-0.38);
      \draw[boson] (0.5,0) -- (0,-0.8);
      \draw[boson] (0,-0.8) -- (0,-1.5);
      \filldraw (-0.5,0) circle (0.05);
      \filldraw (0.5,0) circle (0.05);
      \filldraw (0,0) circle (0.05);
      \filldraw (0,-1.5) circle (0.05);
    \end{tikzpicture}
    \caption{1/2}
  \end{subfigure}
   \begin{subfigure}[b]{0.1\textwidth}
    \begin{tikzpicture}
      [scale=1]
      \draw[boson] (-0.5,0) -- (-0.2,-0.5);
      \draw[boson] (-0.2,-0.5) -- (0,-0.8);
      \draw[boson] (0,0) -- (-0.29,-0.38);
      \draw[scalar] (0.5,0) -- (0,-0.8);
      \draw[scalar] (0,-0.8) -- (0,-1.5);
      \filldraw (-0.5,0) circle (0.05);
      \filldraw (0.5,0) circle (0.05);
      \filldraw (0,0) circle (0.05);
      \filldraw (0,-1.5) circle (0.05);
    \end{tikzpicture}
    \caption{1/2}
    \end{subfigure}
    \caption{Feynman diagrams due to two-scalar two-graviton bulk vertices.}
    \label{fig:bulk2scalar}
  \end{figure}

  \clearpage
\bibliography{bibpaper}

\end{document}